\documentclass[prd,aps,showpacs,notitlepage,nofootinbib]{revtex4-1}

\usepackage{verbatim}
\usepackage{epsfig}
\usepackage{amssymb}
\usepackage{amsmath}
\usepackage{bm}
\usepackage{color}

\begin{document}
\title{Microwave Background Correlations from Dipole Anisotropy Modulation}

\date{\today}
\author{Simone Aiola}
\email{sia21@pitt.edu}
\author{Bingjie Wang}
\author{Arthur Kosowsky}
\affiliation{Department of Physics and
Astronomy, University of Pittsburgh, 3941 O'Hara Street, Pittsburgh, PA 15260 USA}
\affiliation{Pittsburgh Particle Physics, Astrophysics, and Cosmology Center, Pittsburgh, PA 15260 USA}

\author{Tina Kahniashvili}
\affiliation{The McWilliams Center for Cosmology and Department of Physics, Carnegie Mellon University, 5000 Forbes Ave., Pittsburgh, PA 15213 USA}
\affiliation{Department of Physics, Laurentian University, Ramsey Lake Road, Sudbury, ON P3E 2C6, Canada} 
\affiliation{Abastumani Astrophysical Observatory, Ilia State University, 3-5 Cholokashvili St. Tbilisi, Georgia, GE-0194}

\author{Hassan Firouzjahi}
\affiliation{School of Astronomy, Institute for Research in Fundamental Sciences (IPM), P.~O.~Box 19395-5531, Tehran, Iran.}

\begin{abstract}
Full-sky maps of the cosmic microwave background temperature reveal a 7\% asymmetry of fluctuation power between two
halves of the sky. A common phenomenological model for this asymmetry is an overall
dipole modulation of statistically isotropic fluctuations, which produces particular off-diagonal
correlations between multipole coefficients. We compute these correlations and construct corresponding
estimators for the amplitude and direction of the dipole modulation. 
Applying these estimators to various cut-sky temperature maps from Planck and WMAP data
shows consistency with a dipole modulation, differing from a null signal at 2.5$\sigma$, with an amplitude
and direction consistent with previous fits based on the temperature fluctuation power. 
The signal is scale dependent, with a statistically significant amplitude at angular scales larger than 2 degrees. 
Future measurements of microwave background polarization and gravitational lensing can increase the significance
of the signal. 
If the signal is not a statistical fluke in an isotropic Universe, it requires new physics beyond the standard model of cosmology.
\end{abstract}

 
\maketitle

\section{Introduction\label{Intro}}
The statistical isotropy of the cosmic microwave background (CMB) at large angular scales has been questioned
since the first data release of the WMAP satellite \cite{Bennett2003}. 
Independent studies performed on different WMAP data releases \cite{Eriksen2004, Hansen2004, Axelsson2013}
show that the microwave temperature sky possesses a hemispherical power asymmetry, 
exhibiting more large-scale power in one half of the sky than the other.
Recently, this finding has been confirmed with a significance greater than $3\sigma$
with CMB temperature maps from the first data release of the Planck experiment \cite{PlanckI}.
The power asymmetry has been detected using multiple techniques, including spatial variation of the 
temperature power spectrum for multipoles up to $l = 600$ \cite{PlanckXXIII}
and measurements of the local variance of the CMB temperature map \cite{Akrami2014, Adhikari2015}.
For $l >600$, the amplitude of the  power asymmetry drops quickly with $l$ \cite{Flender13, Adhikari2015}.

A phenomenological model for the hemispherical power asymmetry is a statistically isotropic sky $\Theta({\bf \hat n})$
times a dipole modulation of the temperature anisotropy amplitude:
\begin{equation}
{\tilde\Theta}({\bf \hat n}) = \left(1 + {\bf\hat n}\cdot{\bf A}\right)\Theta({\bf\hat n}),
\label{dipole_modulation}
\end{equation}
where the vector ${\bf A}$ gives the dipole amplitude and sky direction of the asymmetry \cite{Gordon2005}.
This phenomenological model has been tested on large scales ($l < 100$) with both WMAP \cite{Eriksen2007, Hoftuft2009} 
and Planck (\cite{PlanckXXIII}, hereafter PLK13) data, showing a  dipole modulation with the amplitude $| {\bf A} | \simeq 0.07$ along the direction $(\ell,b) \simeq (220^{\circ}, -20^{\circ})$ in galactic coordinates,
with a significance at a level $\geq 3\sigma$. 
Further analysis at intermediate scales $100 < l < 600$ shows that the amplitude of the dipole modulation is also scale dependent \cite{Hanson2009}. 

If a dipole modulation in the form of Eq.~(\ref{dipole_modulation}) is present, it induces off-diagonal correlations between multipole components with differing $l$ values. 
Similar techniques have been employed to study both the dipole modulation \cite{Prunet2005, Hanson2009, Moss2011, Rath2014} and the local peculiar velocity 
\cite{Kosowsky2011, Amendola2011, PlanckXXVII, Jeong2013}.
In this work, we exploit these correlations to construct estimators for the Cartesian components of the vector ${\bf A}$ as function of the multipole. 
These estimators are then applied to publicly available foreground-cleaned Planck CMB temperature maps. 
We constrain the scale dependence over a multipole range $2 \leq l \leq 600$, as well as determine the statistical significance of the observed geometrical configuration as function of the multipole.
Throughout this analysis, we adopt realistic masking of the galactic contamination. We test our findings against possible instrumental systematics and residual foregrounds.

In the following Section, we derive estimators for the dipole modulation components and their variances for a cosmic-variance limited CMB temperature map. 
Section~\ref{Pipeline} presents and tests a pipeline for deriving these estimators from observed maps, 
showing how to correct for partial sky coverage. Using simulated CMB maps, we estimate the covariance matrix of the components of the dipole vector, as well as testing for possible systematic effects. 
Section~\ref{Data} describes the Planck temperature data we use to obtain the results
in Sec.~\ref{Results}. We estimate the components of the dipole modulation vector and assess their
statistical significance, finding departures from zero at the 2 to 3 $\sigma$ level. 
The best-fit dipole modulation signal is an unexpectedly good fit to the data, suggesting that we have neglected additional correlations
in modeling the temperature sky. We also perform a Monte Carlo analysis to estimate how the dipole
modulation depends on angular scale, confirming previous work showing the power modulation becoming
undetectable for angular scales less than $0.4^\circ$. 
Finally, Sec.~\ref{Conclusions} gives a discussion of the significance of the results and possible implications for models of primordial perturbations.

\section{Dipole-Modulation-Induced Correlations and Estimators\label{Correlations}}

Assuming the phenomenological model described by Eq.~(\ref{dipole_modulation}), 
the dipole dependence on direction can be expressed in terms of the $l=1$ spherical harmonics as
\begin{equation}
{\bf\hat n}\cdot{\bf A} = 2\sqrt{\frac{\pi}{3}}\bigl(A_+ Y_{1-1}({\bf\hat n}) - A_- Y_{1+1}({\bf\hat n}) + A_zY_{10}({\bf\hat n})\bigr)
\label{dipole_Y}
\end{equation}
with the abbreviation $A_{\pm}\equiv (A_x\pm iA_y)/\sqrt{2}$. 
Expanding the
temperature distributions in the usual spherical harmonics,
\begin{equation}
\Theta({\bf\hat n}) = \sum_{lm} a_{lm} Y_{lm}({\bf\hat n}), \qquad\qquad 
{\tilde\Theta}({\bf\hat n}) = \sum_{lm} {\tilde a}_{lm} Y_{lm}({\bf\hat n}),
\label{alm_defs}
\end{equation}
with the usual isotropic expectation values 
\begin{equation}
\left\langle a_{lm}^* a_{l'm'}\right\rangle = C_l \delta_{ll'}\delta_{mm'}.
\label{aa}
\end{equation}
The coefficients must satisfy $a^*_{lm} = (-1)^m a_{l-m}$ and ${\tilde a}^*_{lm} = (-1)^m {\tilde a}_{l-m}$ because the temperature field is real.
The asymmetric multipoles can be expressed in terms of the symmetric 
multipoles as
\begin{equation}
{\tilde a}_{lm} =  a_{lm} +  2\sqrt{\frac{\pi}{3}}\sum_{l'm'}a_{l'm'}(-1)^m \int d{\bf \hat n}\, Y_{l-m}({\bf\hat n})Y_{l'm'} ({\bf\hat n})
\bigl[A_+ Y_{1-1}({\bf\hat n}) - A_- Y_{1+1}({\bf\hat n}) + A_z Y_{10}({\bf\hat n})\bigr].
\label{alm_1}
\end{equation}
The integrals can be performed in terms of the Wigner 3j symbols using the usual Gaunt formula,
\begin{equation}
\int d{\bf\hat n} Y_{l_1m_1}({\bf\hat n})Y_{l_2 m_2}({\bf\hat n})Y_{l_3m_3}({\bf\hat n}) = \sqrt{\frac{(2l_1+1)(2l_2+1)(2l_3+1)}{4\pi}}\begin{pmatrix} l_1 & l_2 & l_3 \\ m_1 & m_2 & m_3 \end{pmatrix}
\begin{pmatrix} l_1 & l_2 & l_3 \\ 0 & 0 & 0 \end{pmatrix}.
\label{gaunt}
\end{equation}
Because $l_3=1$ for each term in Eq.~(\ref{alm_1}), the triangle inequalities obeyed by the 3j symbols give that the only nonzero terms in Eq.~(\ref{alm_1})
are $l'=l$ or $l'=l\pm 1$. For these simple cases, the 3j symbols can be evaluated explicitly. Then it is straightforward to derive
\begin{eqnarray}
\left\langle {\tilde a}^*_{l+1\,m\pm 1} {\tilde a}_{lm}\right\rangle &=& \mp \frac{1}{\sqrt{2}}A_\pm\left(C_l + C_{l+1}\right)
\sqrt{\frac{(l\pm m+2)(l\pm m+1)}{(2l+3)(2l+1)}},\label{aa_pm}\\
\left\langle {\tilde a}^*_{l+1\,m} {\tilde a}_{lm}\right\rangle &=& A_z \left(C_l + C_{l+1}\right)
\sqrt{\frac{(l-m+1)(l+m+1)}{(2l+3)(2l+1)}}.\label{aa_0}
\end{eqnarray}
These off-diagonal correlations between multipole coefficients with different $l$ values are zero for an isotropic
sky. This result was previously found by \cite{Prunet2005}, and represents a special case of the 
bipolar spherical harmonic formalism \cite{Hajian2003}.

It is now simple to construct estimators for the components of ${\bf A}$ from products of multipole coefficients
in a map. Using $A_x = \sqrt{2}{\rm Re}A_+$ and $A_y = \sqrt{2}{\rm Im}A_+$, we obtain the following estimators:
\begin{eqnarray}
\left[A_x\right]_{lm} &\simeq& \frac{-2}{C_l + C_{l+1}}\sqrt{\frac{(2l+3)(2l+1)}{(l+m+2)(l+m+1)}}
\left({\rm Re}\,{\tilde a}_{l+1\,m+1}{\rm Re}\,{\tilde a}_{lm} + {\rm Im}\,{\tilde a}_{l+1\,m+1}{\rm Im}\,{\tilde a}_{lm}\right),\label{ax_est}\\
\left[A_y\right]_{lm} &\simeq& \frac{-2}{C_l + C_{l+1}}\sqrt{\frac{(2l+3)(2l+1)}{(l+m+2)(l+m+1)}}
\left({\rm Re}\,{\tilde a}_{l+1\,m+1}{\rm Im}\,{\tilde a}_{lm} - {\rm Im}\,{\tilde a}_{l+1\,m+1}{\rm Re}\,{\tilde a}_{lm}\right),\label{ay_est}\\
\left[A_z\right]_{lm} &\simeq& \frac{1}{C_l + C_{l+1}}\sqrt{\frac{(2l+3)(2l+1)}{(l+m+1)(l-m+1)}}
\left({\rm Re}\,{\tilde a}_{l+1\,m}{\rm Re}\,{\tilde a}_{lm} + {\rm Im}\,{\tilde a}_{l+1\,m}{\rm Im}\,{\tilde a}_{lm}\right).\label{az_est}
\end{eqnarray}
where $\tilde{a}_{lm}$'s are calculated from a given (real or simulated) map and $C_{l}$'s are estimated from the harmonic coefficients of the 
isotropic map $C_{l} = {(2 l +1)}^{-1} \sum |a_{lm}|^2$. We argue that for small values of the dipole vector $\mathbf{A}$ and (more important)
for a nearly full-sky map $\sum |\tilde{a}_{lm}|^2 \rightarrow \sum |{a}_{lm}|^2$. 
This assumption has been tested for the kinematic dipole modulation induced in the CMB due to our proper motion, showing that
the bias on the estimated power spectrum is much smaller than the cosmic variance error for nearly full-sky surveys \cite{Jeong2013}.
Such estimators, derived under the constraint of constant dipole modulation, can be safely used for the general case 
of a scale-dependent dipole vector ${\bf A}$ by assuming that ${\bf{A}} (l) \simeq {\bf{A}} (l+1)$.
This requirement is trivially satisfied by a small and monotonically decreasing function ${\bf {A}} (l)$.

To compute the variance of these estimators, assume a full-sky microwave background map which is dominated by
cosmic variance; the Planck maps are a good approximation to this ideal. Then $a_{lm}$ is a Gaussian random variable
with variance $\sigma_l^2 = C_l$.  The real and
imaginary parts are also each Gaussian distributed, with a variance half as large. The product 
$x= {\rm Re}\,{\tilde a}_{l+1\,m+1}{\rm Re}\,{\tilde a}_{lm}$, for example, will then have a product normal distribution
with probability density 
\begin{equation}
P(x) = \frac{2}{\pi\sigma_l \sigma_{l+1}} K_0\left(\frac{2|x|}{\sigma_l\sigma_{l+1}}\right)
\label{product_normal}
\end{equation}
with variance $\sigma_x^2 = \sigma_l^2\sigma_{l+1}^2/4$, where $K_0(x)$ is a modified Bessel
function. By the central limit theorem, a sum of random variables with different variances will tend to
a normal distribution with variance equal to the sum of the variances of the random variables; in practice
the sum of two random variables each with a product normal distribution
will be very close to normally distributed, as can be verified numerically
from Eq.~(\ref{product_normal}). Therefore we can treat the sums
of pairs of ${\tilde a}_{lm}$ values in Eqs.~(\ref{ax_est}) to (\ref{az_est}) as normal variables with
variance $\sigma_l^2\sigma_{l+1}^2/2$, and obtain the standard errors for the estimators as
\begin{eqnarray}
\left[\sigma_x\right]_{lm} = \left[\sigma_y\right]_{lm} \simeq \sqrt{\frac{(2l+3)(2l+1)}{2(l+m+2)(l+m+1)}},\label{sigmax_est}\\
\left[\sigma_z\right]_{lm} \simeq \frac{1}{2}\sqrt{\frac{(2l+3)(2l+1)}{2(l+m+1)(l-m+1)}},\label{sigmaz_est}
\end{eqnarray}
with the approximation $C_{l+1}\simeq C_l$. 

For a sky map with cosmic variance, each estimator of the components of ${\bf A}$ for a given value of $l$ and $m$ will
have a low signal-to-noise ratio. Averaging the estimators with inverse variance weighting will give the highest signal-to-noise
ratio. Consider such an estimator for a component of ${\bf A}$
which average all of the multipole moments between $l=a$ and $l=b$:
\begin{eqnarray}
\left[A_x\right] \equiv \sigma_x^2 \sum_{l=a}^{b} \sum_{m=-l}^l \frac{\left[A_x\right]_{lm}}{\left[\sigma_x\right]_{lm}^2},
\label{ax_est_sum}\\
\left[A_y\right] \equiv \sigma_y^2 \sum_{l=a}^{b} \sum_{m=-l}^l \frac{\left[A_y\right]_{lm}}{\left[\sigma_y\right]_{lm}^2},
\label{ay_est_sum}\\
\left[A_z\right] \equiv \sigma_z^2 \sum_{l=a}^{b} \sum_{m=0}^l \frac{\left[A_z\right]_{lm}}{\left[\sigma_z\right]_{lm}^2},
\label{az_est_sum}
\end{eqnarray}
which have standard errors of
\begin{eqnarray}
\sigma_x &=& \sigma_y \equiv \left[\sum_{l=a}^{b} \sum_{m=-l}^l \left[\sigma_x\right]_{lm}^{-2}\right]^{-1/2}  = \left[\frac{2}{3}(b+a+2)(b-a+1)\right]^{-1/2},\label{sigmax}\\
\sigma_z &\equiv& \left[\sum_{l=a}^{b} \sum_{m=0}^l \left[\sigma_z\right]_{lm}^{-2}\right]^{-1/2}
= \left[\frac{4(b-a+1)\left[a(2b+3)(a+b+4) + (b+2)(b+3)\right]}{3(2a+1)(2b+3)}\right]^{-1/2}.\label{sigmaz}
\end{eqnarray}
The sum over $m$ for the $z$ estimator and error runs from $0$ instead of $-l$ because $[A_z]_{l-m} = [A_z]_{lm}$, but the values are distinct for the $x$ and $y$ estimators.

While the Cartesian components are real Gaussian random variables, such that for isotropic models $\langle[A_x] \rangle = \langle[A_y] \rangle = \langle[A_z] \rangle = 0$, 
the amplitude of $\mathbf{A}$ is not Gaussian distributed. Instead, it is described by a chi-square distribution with 3 degrees of freedom, 
which implies $\langle |\mathbf{A}|^2 \rangle \neq 0$ and $p(|\mathbf{A}|^2=0) = 0$, even for an isotropic sky. For this reason,
we consider the properties of the dipole vector $\mathbf{A}$ as function of the multipole, considering each Cartesian component separately. 

\section{Simulations and Analysis Pipeline \label{Pipeline}}
The estimators in Eq.~(\ref{ax_est})-(\ref{az_est}) are clearly unbiased for the case of full-sky CMB map.
However, residual foreground contaminations along the galactic plane as well as point sources may cause a spurious dipole modulation signal, 
which can be interpreted as cosmological. Such highly contaminated regions can be masked out, at the cost of breaking the statistical isotropy of 
the CMB field and inducing off-diagonal correlations between different modes. The effect of the mask, which has a known structure, can be characterized 
and removed. 

\subsection{Characterization of the Mask\label{Mask}}
For a masked sky the original $a_{lm}$ are replaced with their masked counterparts:
\begin{equation}
\overline{a}_{lm} = \int {\bf d} \Omega \Theta({\bf\hat n})W({\bf\hat n}) Y^*_{lm}
\label{masked_alm}
\end{equation}
where $W({\bf\hat n})$ is the mask, with $0 \leq W({\bf\hat n}) \leq 1$.
In this case, Eq.~(\ref{aa}) does not hold, meaning that even for a statistical isotropic but masked sky the estimators in Eq.~(\ref{ax_est})-(\ref{az_est})
will have an expectation value different from zero. This constitutes a bias factor in our estimation of the dipole modulation.

If we expand Eqs.~(\ref{ax_est})--(\ref{az_est}) using the definition of the harmonic coefficients in Eq.~(\ref{alm_1}), it is clear that if a primordial dipole modulation is present, 
the mask transfers power between different Cartesian components. Under the previous assumption ${\bf A} (l) \simeq {\bf A} (l+1)$, the Cartesian components $i,j = x,y,z$ of the dipole 
vector can be written as
\begin{equation}
[\overline{A}_j]_{lm} = \mathbf{\Lambda}_{ji, lm} A_{i,l} + {M}_{j,lm}
\label{lin_comb}
\end{equation}
where $[\overline{A_j}]_{lm}$ is the estimated dipole vector for the masked map, and $\mathbf{\Lambda}_{ji, lm}$ and ${M}_{j,lm}$ are Gaussian random numbers 
determined by the ${\overline{a}}_{lm}$, so dependent only on the geometry of the mask. 
For unmasked skies, these two quantities satisfy $\langle \mathbf{\Lambda}_{ji, lm} \rangle = \delta_{ij} $ and $\langle {M}_{j,lm} \rangle = 0$, ensuring that 
the expectation value of our estimator converges to the true value.

Using Eq.~(\ref{lin_comb}), we can define a transformation to recover the true binned  dipole vector from a masked map:
\begin{equation}
[{A}_i] = \mathbf{\Lambda}^{-1}_{ji} ([\overline{A}_j] - {M}_{j})
\label{deconv}
\end{equation}
where $[\overline{A}_j]$ is the binned dipole vector estimated from a map, and $\mathbf{\Lambda}_{ji}$ and ${M}_{j}$ are the expectation values of $\mathbf{\Lambda}_{ji, lm}$ and ${M}_{j,lm}$, 
binned using the prescription in Eqs.~(\ref{ax_est_sum})--(\ref{az_est_sum}).
For each Cartesian component we divide the multipole range in $19$ bins with uneven spacing, 
$\Delta l = 10$ for $2 \leq l \leq 100$, $\Delta l = 100$ for $101 \leq l \leq 1000$.
For a given mask, the matrix $\mathbf{\Lambda}_{ji}$ and the vector ${M}_{j}$ can be computed by using simulations of isotropic
masked skies. We use an ensemble of $2000$ simulations, and we adopt the apodized \verb|Planck U73| mask, 
following the procedure adopted by PLK13 for the hemispherical power asymmetry analysis. 
For the rest of this work, all estimates of the dipole vector are corrected for the effect of the mask using Eq.~(\ref{deconv}).

\subsection{Simulated Skies\label{Cov_sec}}
We generate $2000$ random masked skies for both isotropic and dipole modulated cases. For the latter, we assume an scale-independent model
with amplitude ${\bf |A|}=0.07$, along the direction in galactic coordinates $(l,b)=(220^{\circ},-20^{\circ})$. 
We adopt a resolution corresponding to the \verb|HEALPix|\footnote{\url{http://healpix.sourceforge.net/}} \cite{HealPix} parameter \verb|NSIDE| = 2048, and we include a Gaussian smoothing of \verb|FWHM| = $5'$ 
to match the resolution of the available maps. The harmonic coefficients ${\tilde a}_{lm}$ are then rescaled by $\sqrt{\tilde{C}_{l}}$, where the power spectrum is calculated 
directly from the masked map. These normalized coefficients (for both isotropic and dipole modulated cases) are then used to estimate the components of the dipole vector.

\begin{figure}[t!]
	\begin{center}
		\includegraphics[width=0.8\textwidth]{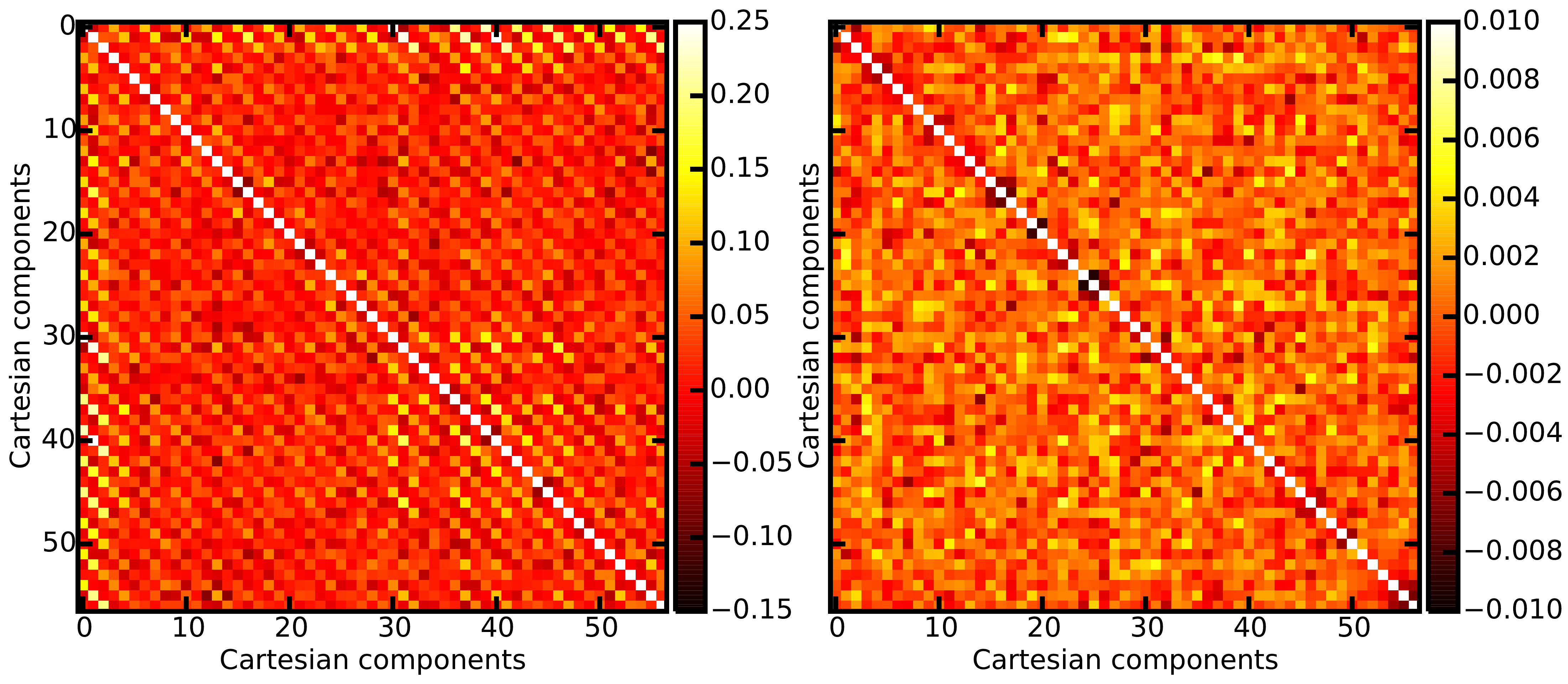}
	\caption{Correlation matrices for the Cartesian components of the dipole vector. These matrices are estimated using $2000$ random simulated skies masked with the apodized Planck U73 mask. 
	The ordering of the components follows the convention defined for the dipole vector. 
	(Left panel) Isotropic skies (${\bf A}$=0). (Right panel) Difference between the correlation matrices for modulated skies, generated using a constant dipole 
	vector across multipoles of magnitude ${\bf |A|}=0.07$ and direction $(\ell,b) = (220^{\circ},-20^{\circ})$, and isotropic skies.}
	\label{Covariance}
	\end{center}
\end{figure}

These simulations also serve the purpose of estimating the covariance matrix ${\bf C}$.
From Eqs.~(\ref{ax_est})-(\ref{az_est}), we expect different Cartesian components to be nearly uncorrelated, even for models with a non-zero dipole modulation, for full-sky maps.
We confirm this numerically with simulations of unmasked skies.
For masked skies, Fig,~(\ref{Covariance}) shows the covariance matrices. The left panel shows the case for isotropic skies
with no dipole modulation. The presence of the mask
induces correlations between multipole bins at scales $100 \lesssim l \lesssim 500$, and also between the largest scales $l \lesssim 40$ with all the other multipole 
bins. However, because of the apodization applied to the mask, the correlation between bins never exceeds $25\%$. 
For comparison, we also show the difference between the correlation matrices for the case of dipole modulated and isotropic skies (right panel).
This is consistent with random noise, which demonstrates that the covariance matrix does not depend significantly
on the amplitude of the dipole modulation.

\subsection{Bias Estimates\label{Tests}}

We determine the mean bias in reconstructing the dipole modulation vector ${\bf A}$ from a masked sky
by computing the mean value of all three Cartesian components reconstructed from $2000$ simulations, for both
isotropic and dipole-modulated skies. 
In both cases, the residual bias vector has components $A_{i} > 0$, with an amplitude of the first bin 
of each Cartesian component below  $6 \times 10^{-3}$.
For the isotropic case, the amplitude of the bias is strongly decreasing with multipole ($|A(l=60)| = 3.8 \times 10^{-4}$), corresponding 
to $0.5\%$ to  $2\%$ of the cosmic variance error for 
the entire multipole range considered. Therefore, the analysis procedure on masked skies 
does not introduce a statistically any significant signal which could be mistaken for dipole modulation.

In the case of dipole-modulated simulations with dipole amplitude $A=0.07$ consistent with PLK13 (see Sec.~\ref{Tests}), 
the amplitude of the bias for each Cartesian component is a constant for all multipoles. 
This indicates that the bias follows the underlying model, and the determination of the scale dependence of the true dipole 
vector will not be affected by such a bias. For this specific case, the amplitude is always $\le 0.8\sigma$ when compared to the cosmic variance, 
specifically $\le 0.1\sigma$ for $l \le 100$. However, this simulated case is unrealistic. 
We do not expect such a big amplitude for the dipole vector at small scales, so the simulated case overestimates the actual bias.

\section{Microwave Sky Data\label{Data}}

We consider a suite of six different foreground-cleaned microwave background temperature maps:\footnote{\url{http://wiki.cosmos.esa.int/planckpla/index.php/Main_Page}}:
\verb|SMICA|, \verb|NILC|, \verb|COMMANDER-Ruler H| and \verb|SEVEM| from the first Planck data release \cite{PlanckXII}, 
and two others processed with the \verb|LGMCA|\footnote{\url{http://www.cosmostat.org/product/lgmca_cmb/}} component separation technique by \cite{Bobin2014}. 
The \verb|LGMCA-PR1| and \verb|LGMCA-WPR1| are based on Planck data only and Planck+WMAP9 data respectively, allowing a non trivial consistency test between these two experiments. 
Each of these maps uses a somewhat different method for separating the microwave background component from foreground emission, 
allowing us to quantify any dependence on the component separation procedure.

Asymmetric beams and inhomogeneous noise may create a systematic dipolar modulation in the sky. 
In order to test this possibility, we analyze the $100$ publicly available \verb|FFP6| single-frequency simulated maps released by the Planck team.
Specifically, we process the simulations for the $100$, $143$ and $217$ GHz channels with our analysis code.
The maximum likelihood analysis shows a bias on small scales, although the values are always less than 0.6 times the cosmic variance for each multipole bin.
Considering only the first $15$ bins ($l_{\rm max} = 600$) gives a result consistent with the isotropic case, with a p-value larger than $0.1$.
The source of the small-scale bias is not yet known, but we simply ignore multipoles $l > 600$ in the present analysis of Planck data.

\begin{figure}[t!]
	\begin{center}
		\includegraphics[width=1.\textwidth]{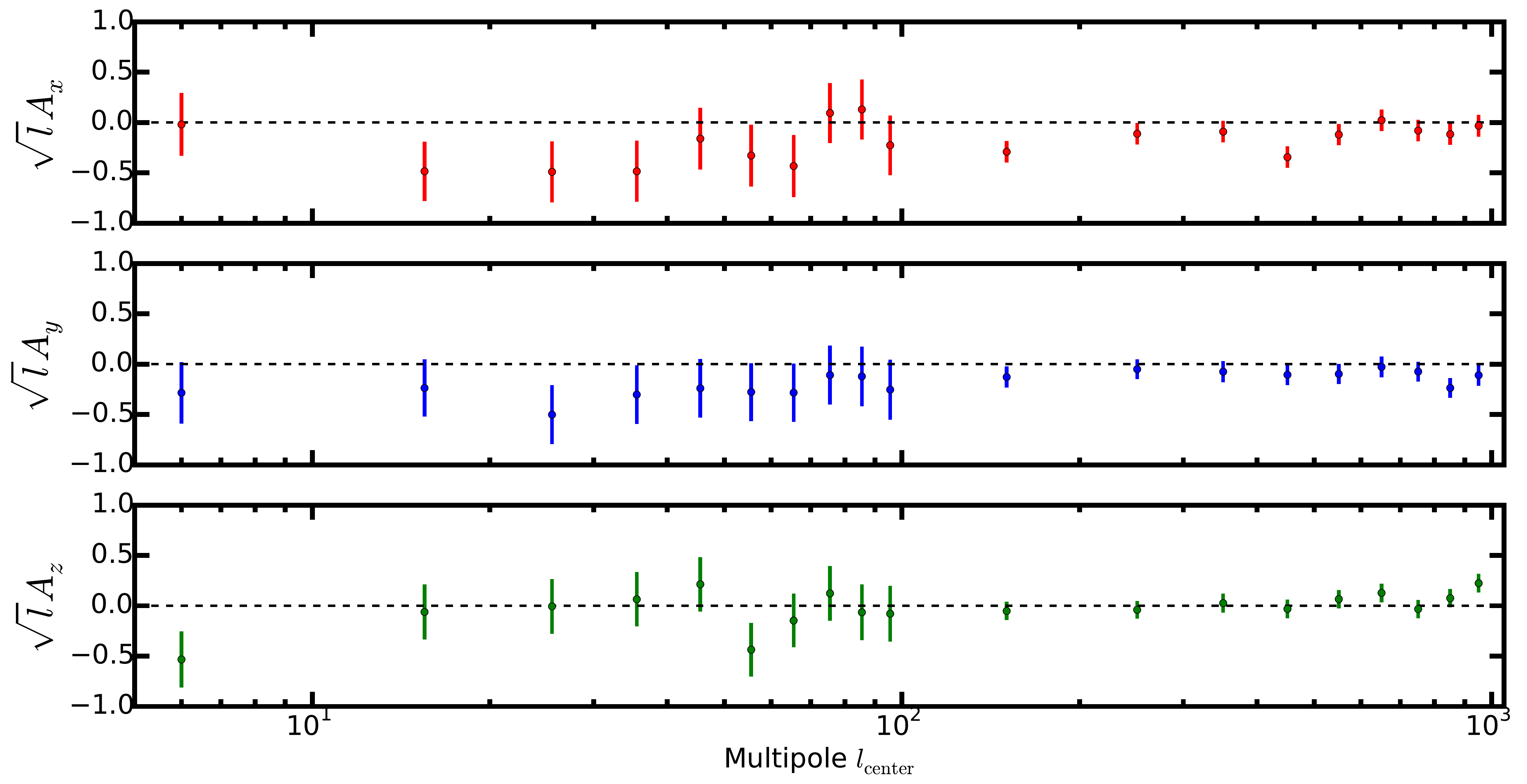}
	\caption{Measured Cartesian components of the dipole vector from the SMICA Planck map as a function of the central
	bin multipole $l_{\rm center}$. The amplitudes are multiplied by $\sqrt{l}$ to enhance visibility of the signal at higher multipoles. 
	The $1\sigma$ errors are the square roots of the covariance matrix diagonal elements. Data at $l > 600$ is not used in our statistical analyses.}
	\label{comp_dipole}
	\end{center}
\end{figure}
\begin{figure}[h!]
	\begin{center}
		\includegraphics[width=1.\textwidth]{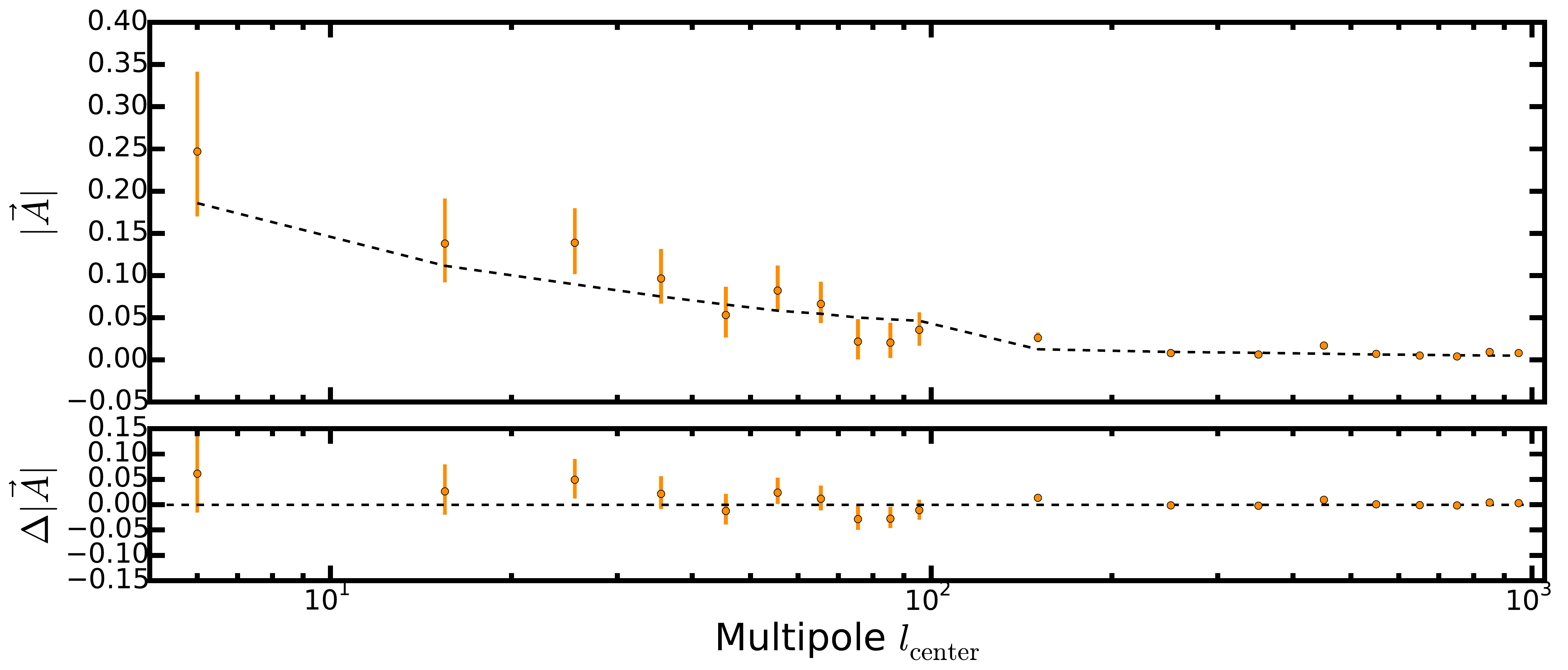}
	\caption{Measured amplitude of the dipole vector from the SMICA Planck map.
	The black dashed line shows the model for the statistical isotropic case ${\bf A}$=0.
	The $1\sigma$ errors are estimated from the $16^{th}$ and $84^{th}$ percentiles of the distribution of the dipole vector amplitudes, calculated from sky simulations processed the same way as the data.}
	\label{amp_dipole}
	\end{center}
\end{figure}

\section{Results\label{Results}}
Figure~(\ref{comp_dipole}) shows the measured values of the Cartesian components of the dipole vector, using the \verb|SMICA| map. 
Similar results are found for the other foreground-cleaned maps, and a direct comparison is shown in Sec.~\ref{GT}. 
Figure~(\ref{amp_dipole}) displays the amplitude of the dipole vector compared with the mean value (black dashed line) obtained from 
isotropic simulations; as pointed out in Sec.~\ref{Correlations}, the expectation value of the amplitude of the dipole vector is different from zero even for the isotropic case.

The data clearly shows two important features:
\begin{itemize}
\item The amplitudes of the components of the estimated dipole vector are decreasing functions of the multipole $l$.
\item The $x$ and $y$ components have a negative sign, which persists over a wide range of multipoles;
the $z$ component is consistent with zero. 
This indicates that the vector is pointing in a sky region $(180^\circ<\ell<270^\circ, b \simeq 0)$, 
in agreement with previous analyses. 
\end{itemize}
We further characterize these basic results in the remainder of this Section.

\subsection{Geometrical Test\label{GT}}
First, we test how likely the observed geometrical configuration of the dipole vector is in an isotropic universe.
To achieve this goal, we need to define a quantity which preserves the information on the direction of the dipole vector (i.e. statistics linear 
in the variables $A_{i}$).  
In addition, the Cartesian components have to be weighted by the cosmic variance, ensuring that our statistics is not dominated 
by the first bins. Therefore, we define the following quantity
\begin{equation}
\alpha = \sum_{i=1}^{3N} (\mathbf{C}^{-1})_{ij} [A]_{j=1,...,3N}
\label{alpha}
\end{equation}
where ($\mathbf{C}^{-1})_{ij}$ are the components of the inverse of the covariance matrix calculated in Sec.~\ref{Cov_sec}, and $[A]_{j=1,...,3N}$
are the three Cartesian components of the binned dipole vector (up to the $N^{\rm th}$ bin) estimated either from a simulated map of measured data.
For an isotropic universe, we expect the three Cartesian components to sum up to zero, such that $\langle \alpha \rangle = 0$ for any choice of $l_{\rm max}$.
This will not be the case if the underlying model is not isotropic (i.e. the expectation values of the Cartesian components are different from zero).
\begin{figure}[t!]
	\begin{center}
		\includegraphics[width=1.\textwidth]{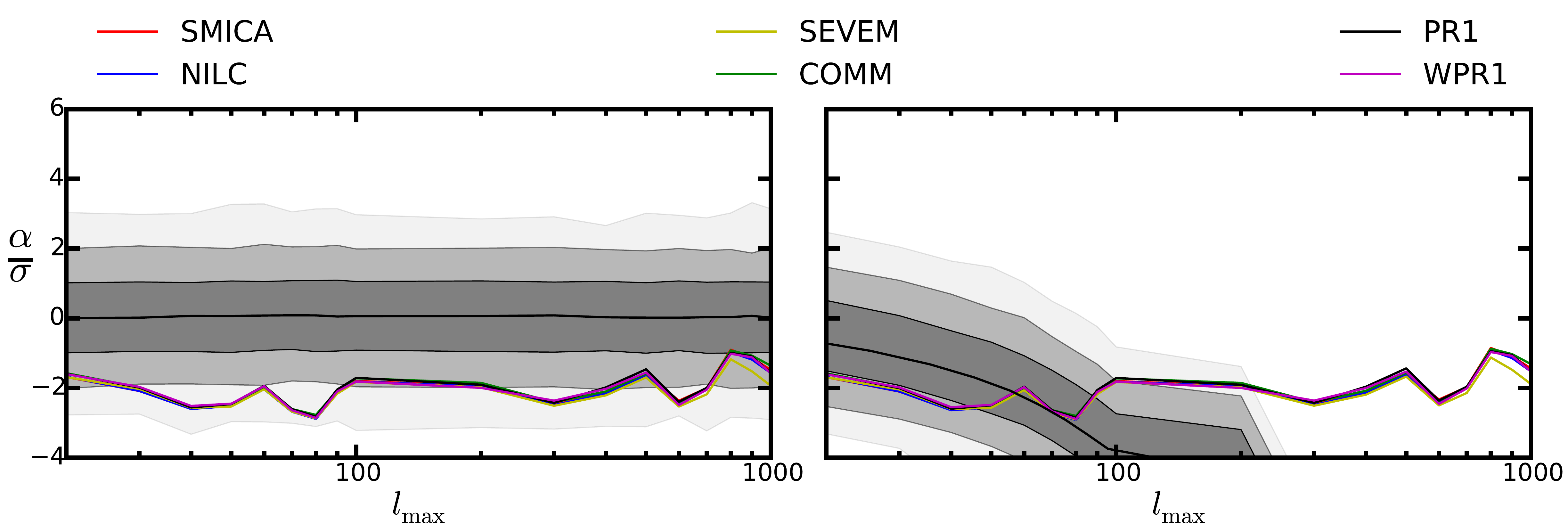}
	\caption{The $\alpha-$parameter from Eq.~(\ref{alpha}), scaled by the standard deviation $\sigma(\alpha)$, as function of the maximum multipole considered $l_{\rm max}$. 
	The colored solid lines are the results from CMB data, showing remarkable consistency between different foreground cleaning methods.
	(Left panel) The shaded bands are estimated using simulations of isotropic masked skies. 
	The distribution $\alpha-$parameter is Gaussian with $\langle \alpha \rangle = 0$. 
	(Right panel) The shaded bands are estimated using simulations of dipole modulated masked skies. 
	The dipole modulation model is $A=0.07$, $(\ell,b) = (220^{\circ}, -20^{\circ})$.
	The confidence regions (colored band) are estimated using percentiles, such that $\pm 1\sigma = [15.87^{th}, 84.13^{th}]$, 
	$\pm 2\sigma = [2.28^{nd}, 97.72^{th}]$ and $\pm 3\sigma = [0.13^{st}, 99.87^{th}]$.}
	\label{Alpha}
	\end{center}
\end{figure}
In Fig.~\ref{Alpha}, we plot the values of the $\alpha$ parameter as function of the maximum multipole considered in the analysis $l_{max}$, 
rescaled by the standard deviation $\sigma(\alpha)$ determined from the simulations of isotropic skies. 
The left panel shows the comparison between the CMB data for all $6$ foreground-cleaned maps, and the simulations for the isotropic case. 
The measured rescaled $\alpha$ parameter has a value that is discrepant from $\alpha=0$ at a level of $2\sigma \lesssim \alpha < 3\sigma$.
This discrepancy is maximized for $l \lesssim 70-80$, which corresponds to what has been previously probed by PLK13.

The right panel of Fig.~\ref{Alpha} compares the measured signal with simulations of dipole modulated skies, using
the covariance matrix $\mathbf{C}$ calculated from the anisotropic simulations.
This test confirms that the signal averaged over multipoles $\lesssim 70$ is consistent with the model proposed by PLK13 (assumed in our anisotropic simulations). 
However, the results are not consistent with a scale-independent dipole modulation, and the amplitude of the dipole modulation vector must be strongly suppressed at higher multipoles. 

\subsection{Model Fitting\label{MCMC}}
Consider a simple power-law model for the dipole modulation defined by $4$ parameters:
\begin{align}
A_{x}^{\rm th} =& A \Big( \frac{l}{60} \Big)^{n} \cos b \cos \ell ,\\
A_{y}^{\rm th} =&A \Big( \frac{l}{60} \Big)^{n} \cos b \sin \ell ,\\
A_{z}^{\rm th}=& -A \Big( \frac{l}{60} \Big)^{n} \sin b,
\end{align}
where $A$ is the amplitude of the dipole vector at the pivot scale of $l = 60$, $n$ is the spectral index of the power law, $b$ is the galactic latitude,
and $\ell$ is the galactic longitude. 
We use a Gaussian likelihood $\mathcal{L}$, such that
\begin{equation}
{\rm ln} \mathcal{L} = -\frac{1}{2} \chi^2 = -\frac{1}{2} ([A_i] - [A_i]^{\rm th})^{\rm T} (\mathbf{C}^{-1})_{ij} ([A_j] - [A_j]^{\rm th})
\end{equation}
where $[A_i]$ are the estimated components from the Planck SMICA map, $[A_i]^{\rm th}$ are the components of the 
assumed model properly binned using Eqs.~(\ref{ax_est_sum})-(\ref{az_est_sum}), and $C$ is the covariance matrix for a dipole-modulated sky displayed in Fig.~\ref{Covariance}.
The parameter space is explored using the Markov chain Monte Carlo sampler \verb|emcee|\footnote{\url{http://dan.iel.fm/emcee/current/}} \cite{emcee}, assuming flat priors  over the ranges $\{A,n\} = \{[0,1], [-2, 2]\}$.
Table~\ref{MCMC_tab}., displays the results for different thresholds of $l_{\rm max}$.

\begin{table}[t!] 
\caption{Best-fit values of the amplitude $A$, spectral index $n$ and direction angles $(\ell,b)$ for the dipole vector, as function of the maximum multipole $l_{\rm max}$. 
The best fit values are corresponds to the $50^{th}$ percentile of the posterior distribution marginalized over the other parameters. 
The errors corresponds to the $16^{th}$ and $84^{th}$ percentiles. For the first case, we consider a model with spectral index 
$n=0$ and $l_{\rm max}=60$, which can be compared with PLK13 findings. Values of the $\chi^2$ corresponding to the best fit model, as well as to 
the isotropic case, are also displayed with corresponding number of degrees of freedom $\nu$.}
\centering  
 \begin{tabular}{c||c|c|c|c||c|c}
 \hline\hline 
$l_{\rm max}$ & $\quad$ $A$  $\quad$ & $\quad$ $n$ $\quad$ & $\quad$ $b[^{\circ}]$ $\quad$ & $\quad$ $\ell[^{\circ}]$ $\quad$ & $\quad$ $\chi^2_{\rm min}(\nu)$ $\quad$ &  $\quad$ $\chi^2_{A=0}(\nu)$ $\quad$\\ [0.5ex]
\hline\hline 
$\quad$ $60$ $\quad$ & $\quad$ $0.063^{+0.028}_{-0.030}$ $\quad$ & $\quad$ $-$ $\quad$ & $\quad$ $-10^{+21}_{-21}$ $\quad$ & $\quad$ $+218^{+24}_{-24}$ $\quad$ & 
$\quad$ 10.3 (15) $\quad$ & $\quad$ 19.5 (18) $\quad$\\
\hline
&&&&&&\\
$\quad$ $200$ $\quad$ & $\quad$ $0.034^{+0.014}_{-0.016}$ $\quad$ & $\quad$ $-0.54^{+0.38}_{-0.22}$ $\quad$ & $\quad$ $-7^{+16}_{-16}$ $\quad$ & $\quad$ $+211^{+19}_{-19}$ $\quad$ & 
$\quad$ 16.4 $(29)\quad$ & $\quad$ 30.8 (33)$\quad$\\
&&&&&&\\
$\quad$ $300$ $\quad$ & $\quad$ $0.029^{+0.012}_{-0.014}$ $\quad$ & $\quad$ $-0.68^{+0.26}_{-0.19}$ $\quad$ & $\quad$ $-11^{+18}_{-16}$ $\quad$ & $\quad$ $+211^{+20}_{-20}$ $\quad$ & 
$\quad$ 18.1 (32)$\quad$ & $\quad$ 31.4 (36)$\quad$\\
&&&&&&\\
$\quad$ $400$ $\quad$ & $\quad$ $0.027^{+0.012}_{-0.014}$ $\quad$ & $\quad$ $-0.74^{+0.22}_{-0.18}$ $\quad$ & $\quad$ $-9^{+19}_{-18}$ $\quad$ & $\quad$ $+212^{+22}_{-21}$ $\quad$ & 
$\quad$ 19.3 (35)$\quad$ & $\quad$ 31.9 (39)$\quad$\\
&&&&&&\\
$\quad$ $500$ $\quad$ & $\quad$ $0.031^{+0.012}_{-0.013}$ $\quad$ & $\quad$ $-0.61^{+0.23}_{-0.15}$ $\quad$ & $\quad$ $-4^{+13}_{-13}$ $\quad$ & $\quad$ $+207^{+16}_{-16}$ $\quad$ & 
$\quad$ 22.5 (38)$\quad$ & $\quad$ 40.2 (42)$\quad$\\
&&&&&&\\
$\quad$ $600$ $\quad$ & $\quad$ $0.031^{+0.011}_{-0.012}$ $\quad$ & $\quad$ $-0.64^{+0.19}_{-0.14}$ $\quad$ & $\quad$ $-1^{+13}_{-14}$ $\quad$ & $\quad$ $+209^{+16}_{-15}$ $\quad$ & 
$\quad$ 23.9 (41)$\quad$ & $\quad$ 41.9 (45)$\quad$\\
\hline\hline
\end{tabular} 
\label{MCMC_tab} 
\end{table}
In the restricted case considering only low multipoles $l<60$ and a flat spectrum $n=0$, our best-fit model
agrees at the $1\sigma$ level with previous analysis by PLK13, for both amplitude and direction.

If $n$ is allowed to vary, 
the amplitude $A$ of the dipole vector at the pivot scale of $60$, as well as the spectral index $n$, are perfectly consistent for three different $l_{\rm max}$ thresholds.
The amplitude is different from the isotropic case $A=0$ at a level of $2\sigma$, and the scale-invariant case $n=0$ with $l_{\rm max} = 400$ is excluded at greater than $3\sigma$ significance.
The value of the galactic longitude $\ell$ is stable to a very high degree, whereas the value of galactic latitude $b$ indicates (although not statistically significant) a migration 
of the pointing from the southern hemisphere to the northern one. 
This is expected due to the effect of the kinematic dipole modulation induced by the proper motion of the solar system with respect to
the microwave background rest frame \cite{Challinor2002, Kosowsky2011}. 
This effect has been detected by Planck \cite{PlanckXXVII}, and results in a dipole modulation in the direction $(\ell,b) = (264^{\circ}, 48^{\circ})$ detectable at high $l$. 

The dipole model is a better fit to the data than isotropic models. Both the Aikake Information Criterion (AIC)
and the Bayes Information Criteron (BIC) \cite{Liddle2007} show sufficient improvement in the fit to justify the addition of
four extra parameters in the model. 
In the specific case of the AIC the dipole model is always favored. 
The improvement is calculated by the relative likelihood of the isotropic model with respect to the dipole modulated case. 
This is defined as ${\rm exp}(({\rm AIC}_{\rm min} - {\rm AIC}_{{\rm A}=0})/2)$, where the AIC factor is corrected for the finite sample size, 
and it corresponds to $0.48,  0.083,  0.13,  0.18 ,  0.013$ and $0.011$ for the models considered in Table~\ref{MCMC_tab}. 
In the case of the BIC, the corresponding values of ${\rm BIC}_{\rm min} - {\rm BIC}_{{\rm A}=0} = -0.5, -0.4,  1.0,  2.1, -2.8$ and $-2.8$. 
The BIC indicates that the dipole modulation is favored only for the cases with $\ell_{\rm max} > 400$, where the parameters are better constrained.
According to \cite{Kass1995}, the improvement even though positive is not strong because $-6 < {\rm BIC}_{\rm min} - {\rm BIC}_{\rm A=0} < -2$.

For the dipole-modulated model, the value of $\chi^2$ is substantially lower than the degrees of freedom. 
This suggests that either the error bars are overestimated, or the data points have correlations which have not been accounted for in the simple dipole model. 
Since the errors are mostly due to cosmic variance on the scales of interest, the error bars cannot have been significantly
overestimated. Therefore, our results may point to additional correlations in the microwave temperature pattern beyond those induced
by a simple dipole modulation of Gaussian random anisotropies. The correlations are unlikely to be due to foregrounds,
since the results show little dependence on different foreground removal techniques.

\section{Conclusions\label{Conclusions}}

The microwave sky seems to exhibit a departure from statistical isotropy, due to half the
sky having slightly more temperature fluctuation power than the other half. 
This work shows that the temperature anisotropies are consistent
with a dipolar amplitude modulation, which induces correlations with multipole coefficients
with $l$ values differing by one. At angular scales of a few degrees and above, the
correlations define a dipole direction which corresponds to the orientation of the
previously known hemispherical power asymmetry, while at smaller scales the direction
migrates to that of the kinematic dipole. Our results show that a dipole modulation
is phenomenologically a good description of the power asymmetry, but that the
modulation must be scale dependent, becoming negligible compared to the
kinematic dipole correlations \cite{Kosowsky2011,PlanckXXVII} on angular scales well below a degree.

The statistical significance of these multipole correlations is between 2$\sigma$ and 3$\sigma$
compared to an isotropic sky, with the error dominated by cosmic variance. The maximum signal
appears at scales $l \lesssim 70$ as seen previously by PLK13. We also find an unusually
low scatter in the dipole component estimates as a function of scale, given the cosmic variance
of an unmodulated Gaussian random field, suggesting that the microwave temperature sky may
have additional correlations not captured by this simple model. 

On the largest scales of the universe, simple models of Inflation predict that the amplitude of any dipole modulation 
due to random perturbations in a statistically isotropic universe should be substantially smaller than that observed.
This departure from statistical isotropy may require new physics in the early Universe. One possible
mechanism is a long-wavelength mode of an additional field which couples to the field
generating perturbations \cite{Erickcek08a, Erickcek08b, Erickcek09, Dai13, Lyth13, Donoghue09, Wang13, Liu13, 
McDonald13, Liddle13, Mazumdar13, DAmico13, Cai13, Chang13, Kohri14, McDonald13b, Kanno13, Liu13b, 
Chang13b, McDonald14, Namjoo13, Akbar13, Firouzjahi14, Namjoo14, Zarei14, Namjoo15, Byrnes15, Kobayashi15, Jazayeri14}.  
If the mode has a wavelength longer than the current Hubble length, an observer
 sees its effect as a gradient. The field gradient modulates background physical quantities such as
 the effective inflaton potential or its slow-roll velocity. The required coupling between long and short
 wavelength modes can be accomplished in the context of squeezed-state non-Gaussianity
 \cite{Namjoo13, Akbar13, Firouzjahi14, Namjoo14}. This mechanism requires a non-trivial scale-dependent non-Gaussianity.
 
Apart from the hypothesis of new physics, foreground contamination and instrumental systematics can 
break the statistical isotropy of the microwave background temperature map.
However, these possibilities can be tested with the available data. Our estimates of the Cartesian components of the dipole vector, 
as functions of angular scale,  are consistent for different foreground-cleaned temperature maps.
The masking adopted in this analysis removes most contaminations from diffuse galactic emission and point sources, 
and our analysis procedure controls possible biases introduced by this procedure.
In addition, realistic instrument simulations provided by the Planck collaboration exclude instrumental effects as
the source of the observed isotropy breaking at the angular scales of interest.
While this work was in preparation, the Planck team has made available the results of a similar analysis
using the 2015 temperature maps \cite{PLK15} (PLK15).
Our estimates of the amplitude and direction of the dipole modulation vector on large-scales ($l_{\max} = 60$) 
are consistent with PLK15 analysis based on Bipolar Spherical Harmonics. The PLK15 analysis does not provide a constraint on the scale-dependency, 
although shows (as for the PLK13 analysis) that the amplitude must decrease at higher multipoles.
PLK15 shows that the amplitude of the dipole vector differs from the isotropic case at a level of $2\sigma-3\sigma$ 
when calculated in cumulative multipole bins $[2,l_{\rm max}]$ for $l_{\rm max}$ up to 320.
This result can be compared with our geometrical test in Fig.~\ref{Alpha}, for which similar results are found.

Additional tests of the dipole modulation will be possible with high-sensitivity
polarization maps covering significant portions of the sky (see, e.g., \cite{AdvACT,PIXIE}). In the standard
inflationary cosmology, microwave polarization and temperature are expected to be only partially 
correlated, giving an additional independent probe of a dipole modulation; a cosmic-variance limited
polarization map will likely double the statistical significance of the signal studied here. Gravitational
lensing of the microwave background over large sky regions provides another nearly independent probe
which will be realized in the near future. We will consider these possibilities elsewhere. If these probes
substantially increase the statistical significance of the dipolar modulation signal, we will be forced into
some significant modification to the inferred physics of the early Universe.

\begin{acknowledgments}
We thank the International Center for Theoretical Physics in Trieste, where this work was initiated. S.A. thanks Federico 
Bianchini and Jeff Newman for insightful discussions. 
This work made use of the GNU Scientific Library, the NIST Digital Library of Mathematical Functions,
the NASA Astrophysical Data System for bibliographic information, and the HEALPix package \cite{HealPix} for map pixelization. 
S.A.\ and A.K.\ are supported by NSF grant 1312380, and T.K.\ is supported by the Swiss NSF SCOPES Grant IZ7370-152581,
the CMU Berkman foundation, the NSF Grants AST-1109180 and the Georgian Shota Rustaveli NSF grant FR/339/6-350/14.
\end{acknowledgments}

\bibliography{dipole_modulation.bib} 

\begin{thebibliography}{10}%
\makeatletter
\providecommand \@ifxundefined [1]{%
 \ifx #1\undefined \expandafter \@firstoftwo
 \else \expandafter \@secondoftwo
\fi
}%
\providecommand \@ifnum [1]{%
 \ifnum #1\expandafter \@firstoftwo
 \else \expandafter \@secondoftwo
\fi
}%
\providecommand \enquote [1]{``#1''}%
\providecommand \bibnamefont  [1]{#1}%
\providecommand \bibfnamefont [1]{#1}%
\providecommand \citenamefont [1]{#1}%
\providecommand\href[0]{\@sanitize\@href}%
\providecommand\@href[1]{\endgroup\@@startlink{#1}\endgroup\@@href}%
\providecommand\@@href[1]{#1\@@endlink}%
\providecommand \@sanitize [0]{\begingroup\catcode`\&12\catcode`\#12\relax}%
\@ifxundefined \pdfoutput {\@firstoftwo}{%
 \@ifnum{\z@=\pdfoutput}{\@firstoftwo}{\@secondoftwo}%
}{%
 \providecommand\@@startlink[1]{\leavevmode\special{html:<a href="#1">}}%
 \providecommand\@@endlink[0]{\special{html:</a>}}%
}{%
 \providecommand\@@startlink[1]{%
  \leavevmode
  \pdfstartlink
   attr{/Border[0 0 1 ]/H/I/C[0 1 1]}%
   user{/Subtype/Link/A<</Type/Action/S/URI/URI(#1)>>}%
  \relax
 }%
 \providecommand\@@endlink[0]{\pdfendlink}%
}%
\providecommand \url  [0]{\begingroup\@sanitize \@url }%
\providecommand \@url [1]{\endgroup\@href {#1}{\urlprefix}}%
\providecommand \urlprefix [0]{URL }%
\providecommand \Eprint[0]{\href }%
\@ifxundefined \urlstyle {%
  \providecommand \doi [1]{doi:\discretionary{}{}{}#1}%
}{%
  \providecommand \doi [0]{doi:\discretionary{}{}{}\begingroup
  \urlstyle{rm}\Url }%
}%
\providecommand \doibase [0]{http://dx.doi.org/}%
\providecommand \Doi[1]{\href{\doibase#1}}%
\providecommand \bibAnnote [3]{%
  \BibitemShut{#1}%
  \begin{quotation}\noindent
    \textsc{Key:}\ #2\\\textsc{Annotation:}\ #3%
  \end{quotation}%
}%
\providecommand \bibAnnoteFile [2]{%
  \IfFileExists{#2}{\bibAnnote {#1} {#2} {\input{#2}}}{}%
}%
\providecommand \typeout [0]{\immediate \write \m@ne }%
\providecommand \selectlanguage [0]{\@gobble}%
\providecommand \bibinfo [0]{\@secondoftwo}%
\providecommand \bibfield [0]{\@secondoftwo}%
\providecommand \translation [1]{[#1]}%
\providecommand \BibitemOpen[0]{}%
\providecommand \bibitemStop [0]{}%
\providecommand \bibitemNoStop [0]{.\EOS\space}%
\providecommand \EOS [0]{\spacefactor3000\relax}%
\providecommand \BibitemShut [1]{\csname bibitem#1\endcsname}%
\bibitem{Bennett2003}%
  \BibitemOpen
  \bibfield{author}{%
  \bibinfo {author} {\bibfnamefont{C.~L.}\ \bibnamefont{{Bennett}}}, \bibinfo
  {author} {\bibfnamefont{M.}~\bibnamefont{{Halpern}}}, \bibinfo {author}
  {\bibfnamefont{G.}~\bibnamefont{{Hinshaw}}}, \bibinfo {author}
  {\bibfnamefont{N.}~\bibnamefont{{Jarosik}}}, \bibinfo {author}
  {\bibfnamefont{A.}~\bibnamefont{{Kogut}}}, \bibinfo {author}
  {\bibfnamefont{M.}~\bibnamefont{{Limon}}}, \bibinfo {author}
  {\bibfnamefont{S.~S.}\ \bibnamefont{{Meyer}}}, \bibinfo {author}
  {\bibfnamefont{L.}~\bibnamefont{{Page}}}, \bibinfo {author}
  {\bibfnamefont{D.~N.}\ \bibnamefont{{Spergel}}}, \bibinfo {author}
  {\bibfnamefont{G.~S.}\ \bibnamefont{{Tucker}}}, \bibinfo {author}
  {\bibfnamefont{E.}~\bibnamefont{{Wollack}}}, \bibinfo {author}
  {\bibfnamefont{E.~L.}\ \bibnamefont{{Wright}}}, \bibinfo {author}
  {\bibfnamefont{C.}~\bibnamefont{{Barnes}}}, \bibinfo {author}
  {\bibfnamefont{M.~R.}\ \bibnamefont{{Greason}}}, \bibinfo {author}
  {\bibfnamefont{R.~S.}\ \bibnamefont{{Hill}}}, \bibinfo {author}
  {\bibfnamefont{E.}~\bibnamefont{{Komatsu}}}, \bibinfo {author}
  {\bibfnamefont{M.~R.}\ \bibnamefont{{Nolta}}}, \bibinfo {author}
  {\bibfnamefont{N.}~\bibnamefont{{Odegard}}}, \bibinfo {author}
  {\bibfnamefont{H.~V.}\ \bibnamefont{{Peiris}}}, \bibinfo {author}
  {\bibfnamefont{L.}~\bibnamefont{{Verde}}},\ and\ \bibinfo {author}
  {\bibfnamefont{J.~L.}\ \bibnamefont{{Weiland}}},\ }%
  \bibfield{journal}{%
  \Doi{10.1086/377253}{\bibinfo {journal} {Astrophys. J. Suppl. Ser.}}\ }%
  \textbf{\bibinfo {volume} {148}},\ \bibinfo {pages} {1} (\bibinfo {year}
  {2003}),\ \Eprint{http://arxiv.org/abs/astro-ph/0302207}{astro-ph/0302207}%
  \bibAnnoteFile{NoStop}{Bennett2003}%
\bibitem{Eriksen2004}%
  \BibitemOpen
  \bibfield{author}{%
  \bibinfo {author} {\bibfnamefont{H.~K.}\ \bibnamefont{{Eriksen}}}, \bibinfo
  {author} {\bibfnamefont{F.~K.}\ \bibnamefont{{Hansen}}}, \bibinfo {author}
  {\bibfnamefont{A.~J.}\ \bibnamefont{{Banday}}}, \bibinfo {author}
  {\bibfnamefont{K.~M.}\ \bibnamefont{{G{\'o}rski}}},\ and\ \bibinfo {author}
  {\bibfnamefont{P.~B.}\ \bibnamefont{{Lilje}}},\ }%
  \bibfield{journal}{%
  \Doi{10.1086/382267}{\bibinfo {journal} {\apj}}\ }%
  \textbf{\bibinfo {volume} {605}},\ \bibinfo {pages} {14} (\bibinfo {year}
  {2004}),\ \Eprint{http://arxiv.org/abs/astro-ph/0307507}{astro-ph/0307507}%
  \bibAnnoteFile{NoStop}{Eriksen2004}%
\bibitem{Hansen2004}%
  \BibitemOpen
  \bibfield{author}{%
  \bibinfo {author} {\bibfnamefont{F.~K.}\ \bibnamefont{{Hansen}}}, \bibinfo
  {author} {\bibfnamefont{A.~J.}\ \bibnamefont{{Banday}}},\ and\ \bibinfo
  {author} {\bibfnamefont{K.~M.}\ \bibnamefont{{G{\'o}rski}}},\ }%
  \bibfield{journal}{%
  \Doi{10.1111/j.1365-2966.2004.08229.x}{\bibinfo {journal} {Mon. Not. R.
  Astron Soc.}}\ }%
  \textbf{\bibinfo {volume} {354}},\ \bibinfo {pages} {641} (\bibinfo {year}
  {2004}),\ \Eprint{http://arxiv.org/abs/astro-ph/0404206}{astro-ph/0404206}%
  \bibAnnoteFile{NoStop}{Hansen2004}%
\bibitem{Axelsson2013}%
  \BibitemOpen
  \bibfield{author}{%
  \bibinfo {author} {\bibfnamefont{M.}~\bibnamefont{{Axelsson}}}, \bibinfo
  {author} {\bibfnamefont{Y.}~\bibnamefont{{Fantaye}}}, \bibinfo {author}
  {\bibfnamefont{F.~K.}\ \bibnamefont{{Hansen}}}, \bibinfo {author}
  {\bibfnamefont{A.~J.}\ \bibnamefont{{Banday}}}, \bibinfo {author}
  {\bibfnamefont{H.~K.}\ \bibnamefont{{Eriksen}}},\ and\ \bibinfo {author}
  {\bibfnamefont{K.~M.}\ \bibnamefont{{Gorski}}},\ }%
  \bibfield{journal}{%
  \Doi{10.1088/2041-8205/773/1/L3}{\bibinfo {journal} {Astrophys. J. Lett.}}\
  }%
  \textbf{\bibinfo {volume} {773}},\ \bibinfo {eid} {L3} (\bibinfo {year}
  {2013}),\ \Eprint{http://arxiv.org/abs/1303.5371}{arXiv:1303.5371}%
  \bibAnnoteFile{NoStop}{Axelsson2013}%
\bibitem{PlanckI}%
  \BibitemOpen
  \bibfield{author}{%
  \bibinfo {author} {\bibnamefont{{Planck Collaboration}}},\ }%
  \bibfield{journal}{%
  \Doi{10.1051/0004-6361/201321529}{\bibinfo {journal} {Astron. Astrophys.}}\
  }%
  \textbf{\bibinfo {volume} {571}},\ \bibinfo {eid} {A1} (\bibinfo {year}
  {2014}),\ \Eprint{http://arxiv.org/abs/1303.5062}{arXiv:1303.5062}%
  \bibAnnoteFile{NoStop}{PlanckI}%
\bibitem{PlanckXXIII}%
  \BibitemOpen
  \bibfield{author}{%
  \bibinfo {author} {\bibnamefont{{Planck Collaboration}}},\ }%
  \bibfield{journal}{%
  \Doi{10.1051/0004-6361/201321534}{\bibinfo {journal} {Astron. Astrophys.}}\
  }%
  \textbf{\bibinfo {volume} {571}},\ \bibinfo {eid} {A23} (\bibinfo {year}
  {2014}),\ \Eprint{http://arxiv.org/abs/1303.5083}{arXiv:1303.5083}%
  \bibAnnoteFile{NoStop}{PlanckXXIII}%
\bibitem{Akrami2014}%
  \BibitemOpen
  \bibfield{author}{%
  \bibinfo {author} {\bibfnamefont{Y.}~\bibnamefont{{Akrami}}}, \bibinfo
  {author} {\bibfnamefont{Y.}~\bibnamefont{{Fantaye}}}, \bibinfo {author}
  {\bibfnamefont{A.}~\bibnamefont{{Shafieloo}}}, \bibinfo {author}
  {\bibfnamefont{H.~K.}\ \bibnamefont{{Eriksen}}}, \bibinfo {author}
  {\bibfnamefont{F.~K.}\ \bibnamefont{{Hansen}}}, \bibinfo {author}
  {\bibfnamefont{A.~J.}\ \bibnamefont{{Banday}}},\ and\ \bibinfo {author}
  {\bibfnamefont{K.~M.}\ \bibnamefont{{G{\'o}rski}}},\ }%
  \bibfield{journal}{%
  \Doi{10.1088/2041-8205/784/2/L42}{\bibinfo {journal} {Astrophys. J. Lett.}}\
  }%
  \textbf{\bibinfo {volume} {784}},\ \bibinfo {eid} {L42} (\bibinfo {year}
  {2014}),\ \Eprint{http://arxiv.org/abs/1402.0870}{arXiv:1402.0870}%
  \bibAnnoteFile{NoStop}{Akrami2014}%
\bibitem{Adhikari2015}%
  \BibitemOpen
  \bibfield{author}{%
  \bibinfo {author} {\bibfnamefont{S.}~\bibnamefont{{Adhikari}}},\ }%
  \bibfield{journal}{%
  \Doi{10.1093/mnras/stu2408}{\bibinfo {journal} {Mon. Not. R. Astron Soc.}}\
  }%
  \textbf{\bibinfo {volume} {446}},\ \bibinfo {pages} {4232} (\bibinfo {year}
  {2015}),\ \Eprint{http://arxiv.org/abs/1408.5396}{arXiv:1408.5396}%
  \bibAnnoteFile{NoStop}{Adhikari2015}%
\bibitem{Flender13}%
  \BibitemOpen
  \bibfield{author}{%
  \bibinfo {author} {\bibfnamefont{S.}~\bibnamefont{{Flender}}}\ and\ \bibinfo
  {author} {\bibfnamefont{S.}~\bibnamefont{{Hotchkiss}}},\ }%
  \bibfield{journal}{%
  \Doi{10.1088/1475-7516/2013/09/033}{\bibinfo {journal} {J. Cosmol. Astropart.
  Phys.}}\ }%
  \textbf{\bibinfo {volume} {9}},\ \bibinfo {eid} {033} (\bibinfo {year}
  {2013}),\ \Eprint{http://arxiv.org/abs/1307.6069}{arXiv:1307.6069}%
  \bibAnnoteFile{NoStop}{Flender13}%
\bibitem{Gordon2005}%
  \BibitemOpen
  \bibfield{author}{%
  \bibinfo {author} {\bibfnamefont{C.}~\bibnamefont{{Gordon}}}, \bibinfo
  {author} {\bibfnamefont{W.}~\bibnamefont{{Hu}}}, \bibinfo {author}
  {\bibfnamefont{D.}~\bibnamefont{{Huterer}}},\ and\ \bibinfo {author}
  {\bibfnamefont{T.}~\bibnamefont{{Crawford}}},\ }%
  \bibfield{journal}{%
  \Doi{10.1103/PhysRevD.72.103002}{\bibinfo {journal} {\prd}}\ }%
  \textbf{\bibinfo {volume} {72}},\ \bibinfo {eid} {103002} (\bibinfo {year}
  {2005}),\ \Eprint{http://arxiv.org/abs/astro-ph/0509301}{astro-ph/0509301}%
  \bibAnnoteFile{NoStop}{Gordon2005}%
\bibitem{Eriksen2007}%
  \BibitemOpen
  \bibfield{author}{%
  \bibinfo {author} {\bibfnamefont{H.~K.}\ \bibnamefont{{Eriksen}}}, \bibinfo
  {author} {\bibfnamefont{A.~J.}\ \bibnamefont{{Banday}}}, \bibinfo {author}
  {\bibfnamefont{K.~M.}\ \bibnamefont{{G{\'o}rski}}}, \bibinfo {author}
  {\bibfnamefont{F.~K.}\ \bibnamefont{{Hansen}}},\ and\ \bibinfo {author}
  {\bibfnamefont{P.~B.}\ \bibnamefont{{Lilje}}},\ }%
  \bibfield{journal}{%
  \Doi{10.1086/518091}{\bibinfo {journal} {Astrophys. J. Lett.}}\ }%
  \textbf{\bibinfo {volume} {660}},\ \bibinfo {pages} {L81} (\bibinfo {year}
  {2007}),\ \Eprint{http://arxiv.org/abs/astro-ph/0701089}{astro-ph/0701089}%
  \bibAnnoteFile{NoStop}{Eriksen2007}%
\bibitem{Hoftuft2009}%
  \BibitemOpen
  \bibfield{author}{%
  \bibinfo {author} {\bibfnamefont{J.}~\bibnamefont{{Hoftuft}}}, \bibinfo
  {author} {\bibfnamefont{H.~K.}\ \bibnamefont{{Eriksen}}}, \bibinfo {author}
  {\bibfnamefont{A.~J.}\ \bibnamefont{{Banday}}}, \bibinfo {author}
  {\bibfnamefont{K.~M.}\ \bibnamefont{{G{\'o}rski}}}, \bibinfo {author}
  {\bibfnamefont{F.~K.}\ \bibnamefont{{Hansen}}},\ and\ \bibinfo {author}
  {\bibfnamefont{P.~B.}\ \bibnamefont{{Lilje}}},\ }%
  \bibfield{journal}{%
  \Doi{10.1088/0004-637X/699/2/985}{\bibinfo {journal} {\apj}}\ }%
  \textbf{\bibinfo {volume} {699}},\ \bibinfo {pages} {985} (\bibinfo {year}
  {2009}),\ \Eprint{http://arxiv.org/abs/0903.1229}{arXiv:0903.1229}%
  \bibAnnoteFile{NoStop}{Hoftuft2009}%
\bibitem{Hanson2009}%
  \BibitemOpen
  \bibfield{author}{%
  \bibinfo {author} {\bibfnamefont{D.}~\bibnamefont{{Hanson}}}\ and\ \bibinfo
  {author} {\bibfnamefont{A.}~\bibnamefont{{Lewis}}},\ }%
  \bibfield{journal}{%
  \Doi{10.1103/PhysRevD.80.063004}{\bibinfo {journal} {\prd}}\ }%
  \textbf{\bibinfo {volume} {80}},\ \bibinfo {eid} {063004} (\bibinfo {year}
  {2009}),\ \Eprint{http://arxiv.org/abs/0908.0963}{arXiv:0908.0963}%
  \bibAnnoteFile{NoStop}{Hanson2009}%
\bibitem{Prunet2005}%
  \BibitemOpen
  \bibfield{author}{%
  \bibinfo {author} {\bibfnamefont{S.}~\bibnamefont{{Prunet}}}, \bibinfo
  {author} {\bibfnamefont{J.-P.}\ \bibnamefont{{Uzan}}}, \bibinfo {author}
  {\bibfnamefont{F.}~\bibnamefont{{Bernardeau}}},\ and\ \bibinfo {author}
  {\bibfnamefont{T.}~\bibnamefont{{Brunier}}},\ }%
  \bibfield{journal}{%
  \Doi{10.1103/PhysRevD.71.083508}{\bibinfo {journal} {\prd}}\ }%
  \textbf{\bibinfo {volume} {71}},\ \bibinfo {eid} {083508} (\bibinfo {year}
  {2005}),\ \Eprint{http://arxiv.org/abs/astro-ph/0406364}{astro-ph/0406364}%
  \bibAnnoteFile{NoStop}{Prunet2005}%
\bibitem{Moss2011}%
  \BibitemOpen
  \bibfield{author}{%
  \bibinfo {author} {\bibfnamefont{A.}~\bibnamefont{Moss}}, \bibinfo {author}
  {\bibfnamefont{D.}~\bibnamefont{Scott}}, \bibinfo {author}
  {\bibfnamefont{J.~P.}\ \bibnamefont{Zibin}},\ and\ \bibinfo {author}
  {\bibfnamefont{R.}~\bibnamefont{Battye}},\ }%
  \bibfield{journal}{%
  \Doi{10.1103/PhysRevD.84.023014}{\bibinfo {journal} {Phys. Rev. D}}\ }%
  \textbf{\bibinfo {volume} {84}},\ \bibinfo {pages} {023014} (\bibinfo {year}
  {2011})%
  \bibAnnoteFile{NoStop}{Moss2011}%
\bibitem{Rath2014}%
  \BibitemOpen
  \bibfield{author}{%
  \bibinfo {author} {\bibfnamefont{P.~K.}\ \bibnamefont{{Rath}}}\ and\ \bibinfo
  {author} {\bibfnamefont{P.}~\bibnamefont{{Jain}}},\ }%
  \bibfield{journal}{%
  \Doi{10.1088/1475-7516/2013/12/014}{\bibinfo {journal} {J. Cosmol. Astropart.
  Phys.}}\ }%
  \textbf{\bibinfo {volume} {12}},\ \bibinfo {eid} {014} (\bibinfo {year}
  {2013}),\ \Eprint{http://arxiv.org/abs/1308.0924}{arXiv:1308.0924}%
  \bibAnnoteFile{NoStop}{Rath2014}%
\bibitem{Kosowsky2011}%
  \BibitemOpen
  \bibfield{author}{%
  \bibinfo {author} {\bibfnamefont{A.}~\bibnamefont{{Kosowsky}}}\ and\ \bibinfo
  {author} {\bibfnamefont{T.}~\bibnamefont{{Kahniashvili}}},\ }%
  \bibfield{journal}{%
  \Doi{10.1103/PhysRevLett.106.191301}{\bibinfo {journal} {\prl}}\ }%
  \textbf{\bibinfo {volume} {106}},\ \bibinfo {eid} {191301} (\bibinfo {year}
  {2011}),\ \Eprint{http://arxiv.org/abs/1007.4539}{arXiv:1007.4539}%
  \bibAnnoteFile{NoStop}{Kosowsky2011}%
\bibitem{Amendola2011}%
  \BibitemOpen
  \bibfield{author}{%
  \bibinfo {author} {\bibfnamefont{L.}~\bibnamefont{{Amendola}}}, \bibinfo
  {author} {\bibfnamefont{R.}~\bibnamefont{{Catena}}}, \bibinfo {author}
  {\bibfnamefont{I.}~\bibnamefont{{Masina}}}, \bibinfo {author}
  {\bibfnamefont{A.}~\bibnamefont{{Notari}}}, \bibinfo {author}
  {\bibfnamefont{M.}~\bibnamefont{{Quartin}}},\ and\ \bibinfo {author}
  {\bibfnamefont{C.}~\bibnamefont{{Quercellini}}},\ }%
  \bibfield{journal}{%
  \Doi{10.1088/1475-7516/2011/07/027}{\bibinfo {journal} {J. Cosmol. Astropart.
  Phys.}}\ }%
  \textbf{\bibinfo {volume} {7}},\ \bibinfo {eid} {027} (\bibinfo {year}
  {2011}),\ \Eprint{http://arxiv.org/abs/1008.1183}{arXiv:1008.1183}%
  \bibAnnoteFile{NoStop}{Amendola2011}%
\bibitem{PlanckXXVII}%
  \BibitemOpen
  \bibfield{author}{%
  \bibinfo {author} {\bibnamefont{{Planck Collaboration}}},\ }%
  \bibfield{journal}{%
  \Doi{10.1051/0004-6361/201321556}{\bibinfo {journal} {Astron. Astrophys.}}\
  }%
  \textbf{\bibinfo {volume} {571}},\ \bibinfo {eid} {A27} (\bibinfo {year}
  {2014}),\ \Eprint{http://arxiv.org/abs/1303.5087}{arXiv:1303.5087}%
  \bibAnnoteFile{NoStop}{PlanckXXVII}%
\bibitem{Jeong2013}%
  \BibitemOpen
  \bibfield{author}{%
  \bibinfo {author} {\bibfnamefont{D.}~\bibnamefont{{Jeong}}}, \bibinfo
  {author} {\bibfnamefont{J.}~\bibnamefont{{Chluba}}}, \bibinfo {author}
  {\bibfnamefont{L.}~\bibnamefont{{Dai}}}, \bibinfo {author}
  {\bibfnamefont{M.}~\bibnamefont{{Kamionkowski}}},\ and\ \bibinfo {author}
  {\bibfnamefont{X.}~\bibnamefont{{Wang}}},\ }%
  \bibfield{journal}{%
  \Doi{10.1103/PhysRevD.89.023003}{\bibinfo {journal} {\prd}}\ }%
  \textbf{\bibinfo {volume} {89}},\ \bibinfo {eid} {023003} (\bibinfo {year}
  {2014}),\ \Eprint{http://arxiv.org/abs/1309.2285}{arXiv:1309.2285}%
  \bibAnnoteFile{NoStop}{Jeong2013}%
\bibitem{Hajian2003}%
  \BibitemOpen
  \bibfield{author}{%
  \bibinfo {author} {\bibfnamefont{A.}~\bibnamefont{{Hajian}}}\ and\ \bibinfo
  {author} {\bibfnamefont{T.}~\bibnamefont{{Souradeep}}},\ }%
  \bibfield{journal}{%
  \Doi{10.1086/379757}{\bibinfo {journal} {Astrophys. J. Lett.}}\ }%
  \textbf{\bibinfo {volume} {597}},\ \bibinfo {pages} {L5} (\bibinfo {year}
  {2003}),\ \Eprint{http://arxiv.org/abs/astro-ph/0308001}{astro-ph/0308001}%
  \bibAnnoteFile{NoStop}{Hajian2003}%
\bibitem{HealPix}%
  \BibitemOpen
  \bibfield{author}{%
  \bibinfo {author} {\bibfnamefont{K.~M.}\ \bibnamefont{{G{\'o}rski}}},
  \bibinfo {author} {\bibfnamefont{E.}~\bibnamefont{{Hivon}}}, \bibinfo
  {author} {\bibfnamefont{A.~J.}\ \bibnamefont{{Banday}}}, \bibinfo {author}
  {\bibfnamefont{B.~D.}\ \bibnamefont{{Wandelt}}}, \bibinfo {author}
  {\bibfnamefont{F.~K.}\ \bibnamefont{{Hansen}}}, \bibinfo {author}
  {\bibfnamefont{M.}~\bibnamefont{{Reinecke}}},\ and\ \bibinfo {author}
  {\bibfnamefont{M.}~\bibnamefont{{Bartelmann}}},\ }%
  \bibfield{journal}{%
  \Doi{10.1086/427976}{\bibinfo {journal} {\apj}}\ }%
  \textbf{\bibinfo {volume} {622}},\ \bibinfo {pages} {759} (\bibinfo {year}
  {2005}),\ \Eprint{http://arxiv.org/abs/astro-ph/0409513}{astro-ph/0409513}%
  \bibAnnoteFile{NoStop}{HealPix}%
\bibitem{PlanckXII}%
  \BibitemOpen
  \bibfield{author}{%
  \bibinfo {author} {\bibnamefont{{Planck Collaboration}}},\ }%
  \bibfield{journal}{%
  \Doi{10.1051/0004-6361/201321580}{\bibinfo {journal} {Astron. Astrophys.}}\
  }%
  \textbf{\bibinfo {volume} {571}},\ \bibinfo {eid} {A12} (\bibinfo {year}
  {2014}),\ \Eprint{http://arxiv.org/abs/1303.5072}{arXiv:1303.5072}%
  \bibAnnoteFile{NoStop}{PlanckXII}%
\bibitem{Bobin2014}%
  \BibitemOpen
  \bibfield{author}{%
  \bibinfo {author} {\bibfnamefont{J.}~\bibnamefont{{Bobin}}}, \bibinfo
  {author} {\bibfnamefont{F.}~\bibnamefont{{Sureau}}}, \bibinfo {author}
  {\bibfnamefont{J.-L.}\ \bibnamefont{{Starck}}}, \bibinfo {author}
  {\bibfnamefont{A.}~\bibnamefont{{Rassat}}},\ and\ \bibinfo {author}
  {\bibfnamefont{P.}~\bibnamefont{{Paykari}}},\ }%
  \bibfield{journal}{%
  \Doi{10.1051/0004-6361/201322372}{\bibinfo {journal} {Astron. Astrophys.}}\
  }%
  \textbf{\bibinfo {volume} {563}},\ \bibinfo {eid} {A105} (\bibinfo {year}
  {2014}),\ \Eprint{http://arxiv.org/abs/1401.6016}{arXiv:1401.6016}%
  \bibAnnoteFile{NoStop}{Bobin2014}%
\bibitem{emcee}%
  \BibitemOpen
  \bibfield{author}{%
  \bibinfo {author} {\bibfnamefont{D.}~\bibnamefont{{Foreman-Mackey}}},
  \bibinfo {author} {\bibfnamefont{D.~W.}\ \bibnamefont{{Hogg}}}, \bibinfo
  {author} {\bibfnamefont{D.}~\bibnamefont{{Lang}}},\ and\ \bibinfo {author}
  {\bibfnamefont{J.}~\bibnamefont{{Goodman}}},\ }%
  \bibfield{journal}{%
  \Doi{10.1086/670067}{\bibinfo {journal} {Publ. Astron. Soc. Pac.}}\ }%
  \textbf{\bibinfo {volume} {125}},\ \bibinfo {pages} {306} (\bibinfo {year}
  {2013}),\ \Eprint{http://arxiv.org/abs/1202.3665}{arXiv:1202.3665}%
  \bibAnnoteFile{NoStop}{emcee}%
\bibitem{Challinor2002}%
  \BibitemOpen
  \bibfield{author}{%
  \bibinfo {author} {\bibfnamefont{A.}~\bibnamefont{{Challinor}}}\ and\
  \bibinfo {author} {\bibfnamefont{F.}~\bibnamefont{{van Leeuwen}}},\ }%
  \bibfield{journal}{%
  \Doi{10.1103/PhysRevD.65.103001}{\bibinfo {journal} {\prd}}\ }%
  \textbf{\bibinfo {volume} {65}},\ \bibinfo {eid} {103001} (\bibinfo {year}
  {2002}),\ \Eprint{http://arxiv.org/abs/astro-ph/0112457}{astro-ph/0112457}%
  \bibAnnoteFile{NoStop}{Challinor2002}%
\bibitem{Liddle2007}%
  \BibitemOpen
  \bibfield{author}{%
  \bibinfo {author} {\bibfnamefont{A.~R.}\ \bibnamefont{{Liddle}}},\ }%
  \bibfield{journal}{%
  \Doi{10.1111/j.1745-3933.2007.00306.x}{\bibinfo {journal} {Mon. Not. R.
  Astron Soc.}}\ }%
  \textbf{\bibinfo {volume} {377}},\ \bibinfo {pages} {L74} (\bibinfo {month}
  {May}\ \bibinfo {year} {2007}),\
  \Eprint{http://arxiv.org/abs/astro-ph/0701113}{astro-ph/0701113}%
  \bibAnnoteFile{NoStop}{Liddle2007}%
\bibitem{Kass1995}%
  \BibitemOpen
  \bibfield{author}{%
  \bibinfo {author} {\bibfnamefont{R.~E.}\ \bibnamefont{Kass}}\ and\ \bibinfo
  {author} {\bibfnamefont{A.~E.}\ \bibnamefont{Raftery}},\ }%
  \bibfield{journal}{%
  \bibinfo {journal} {Journal of the American Statistical Association}\ }%
  \textbf{\bibinfo {volume} {90}},\ \bibinfo {pages} {pp. 773} (\bibinfo {year}
  {1995}),\ ISSN \bibinfo {issn} {01621459}%
  \bibAnnoteFile{NoStop}{Kass1995}%
\bibitem{Erickcek08a}%
  \BibitemOpen
  \bibfield{author}{%
  \bibinfo {author} {\bibfnamefont{A.~L.}\ \bibnamefont{{Erickcek}}}, \bibinfo
  {author} {\bibfnamefont{M.}~\bibnamefont{{Kamionkowski}}},\ and\ \bibinfo
  {author} {\bibfnamefont{S.~M.}\ \bibnamefont{{Carroll}}},\ }%
  \bibfield{journal}{%
  \Doi{10.1103/PhysRevD.78.123520}{\bibinfo {journal} {\prd}}\ }%
  \textbf{\bibinfo {volume} {78}},\ \bibinfo {eid} {123520} (\bibinfo {year}
  {2008}),\ \Eprint{http://arxiv.org/abs/0806.0377}{arXiv:0806.0377}%
  \bibAnnoteFile{NoStop}{Erickcek08a}%
\bibitem{Erickcek08b}%
  \BibitemOpen
  \bibfield{author}{%
  \bibinfo {author} {\bibfnamefont{A.~L.}\ \bibnamefont{{Erickcek}}}, \bibinfo
  {author} {\bibfnamefont{S.~M.}\ \bibnamefont{{Carroll}}},\ and\ \bibinfo
  {author} {\bibfnamefont{M.}~\bibnamefont{{Kamionkowski}}},\ }%
  \bibfield{journal}{%
  \Doi{10.1103/PhysRevD.78.083012}{\bibinfo {journal} {\prd}}\ }%
  \textbf{\bibinfo {volume} {78}},\ \bibinfo {eid} {083012} (\bibinfo {year}
  {2008}),\ \Eprint{http://arxiv.org/abs/0808.1570}{arXiv:0808.1570}%
  \bibAnnoteFile{NoStop}{Erickcek08b}%
\bibitem{Erickcek09}%
  \BibitemOpen
  \bibfield{author}{%
  \bibinfo {author} {\bibfnamefont{A.~L.}\ \bibnamefont{{Erickcek}}}, \bibinfo
  {author} {\bibfnamefont{C.~M.}\ \bibnamefont{{Hirata}}},\ and\ \bibinfo
  {author} {\bibfnamefont{M.}~\bibnamefont{{Kamionkowski}}},\ }%
  \bibfield{journal}{%
  \Doi{10.1103/PhysRevD.80.083507}{\bibinfo {journal} {\prd}}\ }%
  \textbf{\bibinfo {volume} {80}},\ \bibinfo {eid} {083507} (\bibinfo {year}
  {2009}),\ \Eprint{http://arxiv.org/abs/0907.0705}{arXiv:0907.0705}%
  \bibAnnoteFile{NoStop}{Erickcek09}%
\bibitem{Dai13}%
  \BibitemOpen
  \bibfield{author}{%
  \bibinfo {author} {\bibfnamefont{L.}~\bibnamefont{{Dai}}}, \bibinfo {author}
  {\bibfnamefont{D.}~\bibnamefont{{Jeong}}}, \bibinfo {author}
  {\bibfnamefont{M.}~\bibnamefont{{Kamionkowski}}},\ and\ \bibinfo {author}
  {\bibfnamefont{J.}~\bibnamefont{{Chluba}}},\ }%
  \bibfield{journal}{%
  \Doi{10.1103/PhysRevD.87.123005}{\bibinfo {journal} {\prd}}\ }%
  \textbf{\bibinfo {volume} {87}},\ \bibinfo {eid} {123005} (\bibinfo {year}
  {2013}),\ \Eprint{http://arxiv.org/abs/1303.6949}{arXiv:1303.6949}%
  \bibAnnoteFile{NoStop}{Dai13}%
\bibitem{Lyth13}%
  \BibitemOpen
  \bibfield{author}{%
  \bibinfo {author} {\bibfnamefont{D.~H.}\ \bibnamefont{{Lyth}}},\ }%
  \bibfield{journal}{%
  \Doi{10.1088/1475-7516/2013/08/007}{\bibinfo {journal} {J. Cosmol. Astropart.
  Phys.}}\ }%
  \textbf{\bibinfo {volume} {8}},\ \bibinfo {eid} {007} (\bibinfo {year}
  {2013}),\ \Eprint{http://arxiv.org/abs/1304.1270}{arXiv:1304.1270}%
  \bibAnnoteFile{NoStop}{Lyth13}%
\bibitem{Donoghue09}%
  \BibitemOpen
  \bibfield{author}{%
  \bibinfo {author} {\bibfnamefont{J.~F.}\ \bibnamefont{{Donoghue}}}, \bibinfo
  {author} {\bibfnamefont{K.}~\bibnamefont{{Dutta}}},\ and\ \bibinfo {author}
  {\bibfnamefont{A.}~\bibnamefont{{Ross}}},\ }%
  \bibfield{journal}{%
  \Doi{10.1103/PhysRevD.80.023526}{\bibinfo {journal} {\prd}}\ }%
  \textbf{\bibinfo {volume} {80}},\ \bibinfo {eid} {023526} (\bibinfo {year}
  {2009}),\ \Eprint{http://arxiv.org/abs/astro-ph/0703455}{astro-ph/0703455}%
  \bibAnnoteFile{NoStop}{Donoghue09}%
\bibitem{Wang13}%
  \BibitemOpen
  \bibfield{author}{%
  \bibinfo {author} {\bibfnamefont{L.}~\bibnamefont{{Wang}}}\ and\ \bibinfo
  {author} {\bibfnamefont{A.}~\bibnamefont{{Mazumdar}}},\ }%
  \bibfield{journal}{%
  \Doi{10.1103/PhysRevD.88.023512}{\bibinfo {journal} {\prd}}\ }%
  \textbf{\bibinfo {volume} {88}},\ \bibinfo {eid} {023512} (\bibinfo {year}
  {2013}),\ \Eprint{http://arxiv.org/abs/1304.6399}{arXiv:1304.6399}%
  \bibAnnoteFile{NoStop}{Wang13}%
\bibitem{Liu13}%
  \BibitemOpen
  \bibfield{author}{%
  \bibinfo {author} {\bibfnamefont{Z.-G.}\ \bibnamefont{{Liu}}}, \bibinfo
  {author} {\bibfnamefont{Z.-K.}\ \bibnamefont{{Guo}}},\ and\ \bibinfo {author}
  {\bibfnamefont{Y.-S.}\ \bibnamefont{{Piao}}},\ }%
  \bibfield{journal}{%
  \Doi{10.1103/PhysRevD.88.063539}{\bibinfo {journal} {\prd}}\ }%
  \textbf{\bibinfo {volume} {88}},\ \bibinfo {eid} {063539} (\bibinfo {year}
  {2013}),\ \Eprint{http://arxiv.org/abs/1304.6527}{arXiv:1304.6527}%
  \bibAnnoteFile{NoStop}{Liu13}%
\bibitem{McDonald13}%
  \BibitemOpen
  \bibfield{author}{%
  \bibinfo {author} {\bibfnamefont{J.}~\bibnamefont{{McDonald}}},\ }%
  \bibfield{journal}{%
  \Doi{10.1088/1475-7516/2013/07/043}{\bibinfo {journal} {J. Cosmol. Astropart.
  Phys.}}\ }%
  \textbf{\bibinfo {volume} {7}},\ \bibinfo {eid} {043} (\bibinfo {year}
  {2013}),\ \Eprint{http://arxiv.org/abs/1305.0525}{arXiv:1305.0525}%
  \bibAnnoteFile{NoStop}{McDonald13}%
\bibitem{Liddle13}%
  \BibitemOpen
  \bibfield{author}{%
  \bibinfo {author} {\bibfnamefont{A.~R.}\ \bibnamefont{{Liddle}}}\ and\
  \bibinfo {author} {\bibfnamefont{M.}~\bibnamefont{{Cort{\^e}s}}},\ }%
  \bibfield{journal}{%
  \Doi{10.1103/PhysRevLett.111.111302}{\bibinfo {journal} {\prl}}\ }%
  \textbf{\bibinfo {volume} {111}},\ \bibinfo {eid} {111302} (\bibinfo {year}
  {2013}),\ \Eprint{http://arxiv.org/abs/1306.5698}{arXiv:1306.5698}%
  \bibAnnoteFile{NoStop}{Liddle13}%
\bibitem{Mazumdar13}%
  \BibitemOpen
  \bibfield{author}{%
  \bibinfo {author} {\bibfnamefont{A.}~\bibnamefont{{Mazumdar}}}\ and\ \bibinfo
  {author} {\bibfnamefont{L.}~\bibnamefont{{Wang}}},\ }%
  \bibfield{journal}{%
  \Doi{10.1088/1475-7516/2013/10/049}{\bibinfo {journal} {J. Cosmol. Astropart.
  Phys.}}\ }%
  \textbf{\bibinfo {volume} {10}},\ \bibinfo {eid} {049} (\bibinfo {year}
  {2013}),\ \Eprint{http://arxiv.org/abs/1306.5736}{arXiv:1306.5736}%
  \bibAnnoteFile{NoStop}{Mazumdar13}%
\bibitem{DAmico13}%
  \BibitemOpen
  \bibfield{author}{%
  \bibinfo {author} {\bibfnamefont{G.}~\bibnamefont{{D'Amico}}}, \bibinfo
  {author} {\bibfnamefont{R.}~\bibnamefont{{Gobbetti}}}, \bibinfo {author}
  {\bibfnamefont{M.}~\bibnamefont{{Kleban}}},\ and\ \bibinfo {author}
  {\bibfnamefont{M.}~\bibnamefont{{Schillo}}},\ }%
  \bibfield{journal}{%
  \Doi{10.1088/1475-7516/2013/11/013}{\bibinfo {journal} {J. Cosmol. Astropart.
  Phys.}}\ }%
  \textbf{\bibinfo {volume} {11}},\ \bibinfo {eid} {013} (\bibinfo {year}
  {2013}),\ \Eprint{http://arxiv.org/abs/1306.6872}{arXiv:1306.6872}%
  \bibAnnoteFile{NoStop}{DAmico13}%
\bibitem{Cai13}%
  \BibitemOpen
  \bibfield{author}{%
  \bibinfo {author} {\bibfnamefont{Y.-F.}\ \bibnamefont{{Cai}}}, \bibinfo
  {author} {\bibfnamefont{W.}~\bibnamefont{{Zhao}}},\ and\ \bibinfo {author}
  {\bibfnamefont{Y.}~\bibnamefont{{Zhang}}},\ }%
  \bibfield{journal}{%
  \Doi{10.1103/PhysRevD.89.023005}{\bibinfo {journal} {\prd}}\ }%
  \textbf{\bibinfo {volume} {89}},\ \bibinfo {eid} {023005} (\bibinfo {year}
  {2014}),\ \Eprint{http://arxiv.org/abs/1307.4090}{arXiv:1307.4090}%
  \bibAnnoteFile{NoStop}{Cai13}%
\bibitem{Chang13}%
  \BibitemOpen
  \bibfield{author}{%
  \bibinfo {author} {\bibfnamefont{Z.}~\bibnamefont{{Chang}}}, \bibinfo
  {author} {\bibfnamefont{X.}~\bibnamefont{{Li}}},\ and\ \bibinfo {author}
  {\bibfnamefont{S.}~\bibnamefont{{Wang}}}}%
   (\bibinfo {year} {2013}),\
  \Eprint{http://arxiv.org/abs/1307.4542}{arXiv:1307.4542}%
  \bibAnnoteFile{NoStop}{Chang13}%
\bibitem{Kohri14}%
  \BibitemOpen
  \bibfield{author}{%
  \bibinfo {author} {\bibfnamefont{K.}~\bibnamefont{{Kohri}}}, \bibinfo
  {author} {\bibfnamefont{C.-M.}\ \bibnamefont{{Lin}}},\ and\ \bibinfo {author}
  {\bibfnamefont{T.}~\bibnamefont{{Matsuda}}},\ }%
  \bibfield{journal}{%
  \Doi{10.1088/1475-7516/2014/08/026}{\bibinfo {journal} {J. Cosmol. Astropart.
  Phys.}}\ }%
  \textbf{\bibinfo {volume} {8}},\ \bibinfo {eid} {026} (\bibinfo {year}
  {2014}),\ \Eprint{http://arxiv.org/abs/1308.5790}{arXiv:1308.5790}%
  \bibAnnoteFile{NoStop}{Kohri14}%
\bibitem{McDonald13b}%
  \BibitemOpen
  \bibfield{author}{%
  \bibinfo {author} {\bibfnamefont{J.}~\bibnamefont{{McDonald}}},\ }%
  \bibfield{journal}{%
  \Doi{10.1088/1475-7516/2013/11/041}{\bibinfo {journal} {J. Cosmol. Astropart.
  Phys.}}\ }%
  \textbf{\bibinfo {volume} {11}},\ \bibinfo {eid} {041} (\bibinfo {year}
  {2013}),\ \Eprint{http://arxiv.org/abs/1309.1122}{arXiv:1309.1122}%
  \bibAnnoteFile{NoStop}{McDonald13b}%
\bibitem{Kanno13}%
  \BibitemOpen
  \bibfield{author}{%
  \bibinfo {author} {\bibfnamefont{S.}~\bibnamefont{{Kanno}}}, \bibinfo
  {author} {\bibfnamefont{M.}~\bibnamefont{{Sasaki}}},\ and\ \bibinfo {author}
  {\bibfnamefont{T.}~\bibnamefont{{Tanaka}}},\ }%
  \bibfield{journal}{%
  \Doi{10.1093/ptep/ptt093}{\bibinfo {journal} {Progress of Theoretical and
  Experimental Physics}}\ }%
  \textbf{\bibinfo {volume} {2013}},\ \bibinfo {eid} {111E01} (\bibinfo {year}
  {2013}),\ \Eprint{http://arxiv.org/abs/1309.1350}{arXiv:1309.1350}%
  \bibAnnoteFile{NoStop}{Kanno13}%
\bibitem{Liu13b}%
  \BibitemOpen
  \bibfield{author}{%
  \bibinfo {author} {\bibfnamefont{Z.-G.}\ \bibnamefont{{Liu}}}, \bibinfo
  {author} {\bibfnamefont{Z.-K.}\ \bibnamefont{{Guo}}},\ and\ \bibinfo {author}
  {\bibfnamefont{Y.-S.}\ \bibnamefont{{Piao}}},\ }%
  \bibfield{journal}{%
  \Doi{10.1140/epjc/s10052-014-3006-0}{\bibinfo {journal} {European Physical
  Journal C}}\ }%
  \textbf{\bibinfo {volume} {74}},\ \bibinfo {eid} {3006} (\bibinfo {year}
  {2014}),\ \Eprint{http://arxiv.org/abs/1311.1599}{arXiv:1311.1599}%
  \bibAnnoteFile{NoStop}{Liu13b}%
\bibitem{Chang13b}%
  \BibitemOpen
  \bibfield{author}{%
  \bibinfo {author} {\bibfnamefont{Z.}~\bibnamefont{{Chang}}}\ and\ \bibinfo
  {author} {\bibfnamefont{S.}~\bibnamefont{{Wang}}}}%
   (\bibinfo {year} {2013}),\
  \Eprint{http://arxiv.org/abs/1312.6575}{arXiv:1312.6575}%
  \bibAnnoteFile{NoStop}{Chang13b}%
\bibitem{McDonald14}%
  \BibitemOpen
  \bibfield{author}{%
  \bibinfo {author} {\bibfnamefont{J.}~\bibnamefont{{McDonald}}},\ }%
  \bibfield{journal}{%
  \Doi{10.1103/PhysRevD.89.127303}{\bibinfo {journal} {\prd}}\ }%
  \textbf{\bibinfo {volume} {89}},\ \bibinfo {eid} {127303} (\bibinfo {year}
  {2014}),\ \Eprint{http://arxiv.org/abs/1403.2076}{arXiv:1403.2076}%
  \bibAnnoteFile{NoStop}{McDonald14}%
\bibitem{Namjoo13}%
  \BibitemOpen
  \bibfield{author}{%
  \bibinfo {author} {\bibfnamefont{M.~H.}\ \bibnamefont{{Namjoo}}}, \bibinfo
  {author} {\bibfnamefont{S.}~\bibnamefont{{Baghram}}},\ and\ \bibinfo {author}
  {\bibfnamefont{H.}~\bibnamefont{{Firouzjahi}}},\ }%
  \bibfield{journal}{%
  \Doi{10.1103/PhysRevD.88.083527}{\bibinfo {journal} {\prd}}\ }%
  \textbf{\bibinfo {volume} {88}},\ \bibinfo {eid} {083527} (\bibinfo {year}
  {2013}),\ \Eprint{http://arxiv.org/abs/1305.0813}{arXiv:1305.0813}%
  \bibAnnoteFile{NoStop}{Namjoo13}%
\bibitem{Akbar13}%
  \BibitemOpen
  \bibfield{author}{%
  \bibinfo {author} {\bibfnamefont{A.}~\bibnamefont{{Akbar Abolhasani}}},
  \bibinfo {author} {\bibfnamefont{S.}~\bibnamefont{{Baghram}}}, \bibinfo
  {author} {\bibfnamefont{H.}~\bibnamefont{{Firouzjahi}}},\ and\ \bibinfo
  {author} {\bibfnamefont{M.~H.}\ \bibnamefont{{Namjoo}}}}%
   (\bibinfo {year} {2013}),\
  \Eprint{http://arxiv.org/abs/1306.6932}{arXiv:1306.6932}%
  \bibAnnoteFile{NoStop}{Akbar13}%
\bibitem{Firouzjahi14}%
  \BibitemOpen
  \bibfield{author}{%
  \bibinfo {author} {\bibfnamefont{H.}~\bibnamefont{{Firouzjahi}}}, \bibinfo
  {author} {\bibfnamefont{J.-O.}\ \bibnamefont{{Gong}}},\ and\ \bibinfo
  {author} {\bibfnamefont{M.~H.}\ \bibnamefont{{Namjoo}}},\ }%
  \bibfield{journal}{%
  \Doi{10.1088/1475-7516/2014/11/037}{\bibinfo {journal} {J. Cosmol. Astropart.
  Phys.}}\ }%
  \textbf{\bibinfo {volume} {11}},\ \bibinfo {eid} {037} (\bibinfo {year}
  {2014}),\ \Eprint{http://arxiv.org/abs/1405.0159}{arXiv:1405.0159}%
  \bibAnnoteFile{NoStop}{Firouzjahi14}%
\bibitem{Namjoo14}%
  \BibitemOpen
  \bibfield{author}{%
  \bibinfo {author} {\bibfnamefont{M.~H.}\ \bibnamefont{{Namjoo}}}, \bibinfo
  {author} {\bibfnamefont{A.}~\bibnamefont{{Akbar Abolhasani}}}, \bibinfo
  {author} {\bibfnamefont{S.}~\bibnamefont{{Baghram}}},\ and\ \bibinfo {author}
  {\bibfnamefont{H.}~\bibnamefont{{Firouzjahi}}},\ }%
  \bibfield{journal}{%
  \Doi{10.1088/1475-7516/2014/08/002}{\bibinfo {journal} {J. Cosmol. Astropart.
  Phys.}}\ }%
  \textbf{\bibinfo {volume} {8}},\ \bibinfo {eid} {002} (\bibinfo {year}
  {2014}),\ \Eprint{http://arxiv.org/abs/1405.7317}{arXiv:1405.7317}%
  \bibAnnoteFile{NoStop}{Namjoo14}%
\bibitem{Zarei14}%
  \BibitemOpen
  \bibfield{author}{%
  \bibinfo {author} {\bibfnamefont{M.}~\bibnamefont{{Zarei}}}}%
   (\bibinfo {year} {2014}),\
  \Eprint{http://arxiv.org/abs/1412.0289}{arXiv:1412.0289 [hep-th]}%
  \bibAnnoteFile{NoStop}{Zarei14}%
\bibitem{Namjoo15}%
  \BibitemOpen
  \bibfield{author}{%
  \bibinfo {author} {\bibfnamefont{M.~H.}\ \bibnamefont{{Namjoo}}}, \bibinfo
  {author} {\bibfnamefont{A.~A.}\ \bibnamefont{{Abolhasani}}}, \bibinfo
  {author} {\bibfnamefont{H.}~\bibnamefont{{Assadullahi}}}, \bibinfo {author}
  {\bibfnamefont{S.}~\bibnamefont{{Baghram}}}, \bibinfo {author}
  {\bibfnamefont{H.}~\bibnamefont{{Firouzjahi}}},\ and\ \bibinfo {author}
  {\bibfnamefont{D.}~\bibnamefont{{Wands}}},\ }%
  \bibfield{journal}{%
  \Doi{10.1088/1475-7516/2015/05/015}{\bibinfo {journal} {J. Cosmol. Astropart.
  Phys.}}\ }%
  \textbf{\bibinfo {volume} {5}},\ \bibinfo {eid} {015} (\bibinfo {year}
  {2015}),\ \Eprint{http://arxiv.org/abs/1411.5312}{arXiv:1411.5312}%
  \bibAnnoteFile{NoStop}{Namjoo15}%
\bibitem{Byrnes15}%
  \BibitemOpen
  \bibfield{author}{%
  \bibinfo {author} {\bibfnamefont{C.~T.}\ \bibnamefont{{Byrnes}}}\ and\
  \bibinfo {author} {\bibfnamefont{E.~R.~M.}\ \bibnamefont{{Tarrant}}}}%
   (\bibinfo {year} {2015}),\
  \Eprint{http://arxiv.org/abs/1502.07339}{arXiv:1502.07339}%
  \bibAnnoteFile{NoStop}{Byrnes15}%
\bibitem{Kobayashi15}%
  \BibitemOpen
  \bibfield{author}{%
  \bibinfo {author} {\bibfnamefont{T.}~\bibnamefont{{Kobayashi}}}, \bibinfo
  {author} {\bibfnamefont{M.}~\bibnamefont{{Cort{\^e}s}}},\ and\ \bibinfo
  {author} {\bibfnamefont{A.~R.}\ \bibnamefont{{Liddle}}},\ }%
  \bibfield{journal}{%
  \Doi{10.1088/1475-7516/2015/05/029}{\bibinfo {journal} {J. Cosmol. Astropart.
  Phys.}}\ }%
  \textbf{\bibinfo {volume} {5}},\ \bibinfo {eid} {029} (\bibinfo {year}
  {2015}),\ \Eprint{http://arxiv.org/abs/1501.05864}{arXiv:1501.05864}%
  \bibAnnoteFile{NoStop}{Kobayashi15}%
\bibitem{Jazayeri14}%
  \BibitemOpen
  \bibfield{author}{%
  \bibinfo {author} {\bibfnamefont{S.}~\bibnamefont{{Jazayeri}}}, \bibinfo
  {author} {\bibfnamefont{Y.}~\bibnamefont{{Akrami}}}, \bibinfo {author}
  {\bibfnamefont{H.}~\bibnamefont{{Firouzjahi}}}, \bibinfo {author}
  {\bibfnamefont{A.~R.}\ \bibnamefont{{Solomon}}},\ and\ \bibinfo {author}
  {\bibfnamefont{Y.}~\bibnamefont{{Wang}}},\ }%
  \bibfield{journal}{%
  \Doi{10.1088/1475-7516/2014/11/044}{\bibinfo {journal} {J. Cosmol. Astropart.
  Phys.}}\ }%
  \textbf{\bibinfo {volume} {11}},\ \bibinfo {eid} {044} (\bibinfo {year}
  {2014}),\ \Eprint{http://arxiv.org/abs/1408.3057}{arXiv:1408.3057}%
  \bibAnnoteFile{NoStop}{Jazayeri14}%
\bibitem{PLK15}%
  \BibitemOpen
  \bibinfo {howpublished}
  {\url{http://www.cosmos.esa.int/documents/387566/522789/Planck_2015_Results_%
XIV_LFI_Isotropy_and_Statistics.pdf/0d58c586-2c28-4ed0-a6bd-e4328dc5cabe}}%
  \bibAnnoteFile{NoStop}{PLK15}%
\bibitem{AdvACT}%
  \BibitemOpen
  \bibfield{author}{%
  \bibinfo {author} {\bibfnamefont{E.}~\bibnamefont{{Calabrese}}}, \bibinfo
  {author} {\bibfnamefont{R.}~\bibnamefont{{Hlo{\v z}ek}}}, \bibinfo {author}
  {\bibfnamefont{N.}~\bibnamefont{{Battaglia}}}, \bibinfo {author}
  {\bibfnamefont{J.~R.}\ \bibnamefont{{Bond}}}, \bibinfo {author}
  {\bibfnamefont{F.}~\bibnamefont{{de Bernardis}}}, \bibinfo {author}
  {\bibfnamefont{M.~J.}\ \bibnamefont{{Devlin}}}, \bibinfo {author}
  {\bibfnamefont{A.}~\bibnamefont{{Hajian}}}, \bibinfo {author}
  {\bibfnamefont{S.}~\bibnamefont{{Henderson}}}, \bibinfo {author}
  {\bibfnamefont{J.~C.}\ \bibnamefont{{Hil}}}, \bibinfo {author}
  {\bibfnamefont{A.}~\bibnamefont{{Kosowsky}}}, \bibinfo {author}
  {\bibfnamefont{T.}~\bibnamefont{{Louis}}}, \bibinfo {author}
  {\bibfnamefont{J.}~\bibnamefont{{McMahon}}}, \bibinfo {author}
  {\bibfnamefont{K.}~\bibnamefont{{Moodley}}}, \bibinfo {author}
  {\bibfnamefont{L.}~\bibnamefont{{Newburgh}}}, \bibinfo {author}
  {\bibfnamefont{M.~D.}\ \bibnamefont{{Niemack}}}, \bibinfo {author}
  {\bibfnamefont{L.~A.}\ \bibnamefont{{Page}}}, \bibinfo {author}
  {\bibfnamefont{B.}~\bibnamefont{{Partridge}}}, \bibinfo {author}
  {\bibfnamefont{N.}~\bibnamefont{{Sehgal}}}, \bibinfo {author}
  {\bibfnamefont{J.~L.}\ \bibnamefont{{Sievers}}}, \bibinfo {author}
  {\bibfnamefont{D.~N.}\ \bibnamefont{{Spergel}}}, \bibinfo {author}
  {\bibfnamefont{S.~T.}\ \bibnamefont{{Staggs}}}, \bibinfo {author}
  {\bibfnamefont{E.~R.}\ \bibnamefont{{Switzer}}}, \bibinfo {author}
  {\bibfnamefont{H.}~\bibnamefont{{Trac}}},\ and\ \bibinfo {author}
  {\bibfnamefont{E.~J.}\ \bibnamefont{{Wollack}}},\ }%
  \bibfield{journal}{%
  \Doi{10.1088/1475-7516/2014/08/010}{\bibinfo {journal} {J. Cosmol. Astropart.
  Phys.}}\ }%
  \textbf{\bibinfo {volume} {8}},\ \bibinfo {eid} {010} (\bibinfo {year}
  {2014}),\ \Eprint{http://arxiv.org/abs/1406.4794}{arXiv:1406.4794}%
  \bibAnnoteFile{NoStop}{AdvACT}%
\bibitem{PIXIE}%
  \BibitemOpen
  \bibfield{author}{%
  \bibinfo {author} {\bibfnamefont{A.}~\bibnamefont{{Kogut}}}, \bibinfo
  {author} {\bibfnamefont{D.~J.}\ \bibnamefont{{Fixsen}}}, \bibinfo {author}
  {\bibfnamefont{D.~T.}\ \bibnamefont{{Chuss}}}, \bibinfo {author}
  {\bibfnamefont{J.}~\bibnamefont{{Dotson}}}, \bibinfo {author}
  {\bibfnamefont{E.}~\bibnamefont{{Dwek}}}, \bibinfo {author}
  {\bibfnamefont{M.}~\bibnamefont{{Halpern}}}, \bibinfo {author}
  {\bibfnamefont{G.~F.}\ \bibnamefont{{Hinshaw}}}, \bibinfo {author}
  {\bibfnamefont{S.~M.}\ \bibnamefont{{Meyer}}}, \bibinfo {author}
  {\bibfnamefont{S.~H.}\ \bibnamefont{{Moseley}}}, \bibinfo {author}
  {\bibfnamefont{M.~D.}\ \bibnamefont{{Seiffert}}}, \bibinfo {author}
  {\bibfnamefont{D.~N.}\ \bibnamefont{{Spergel}}},\ and\ \bibinfo {author}
  {\bibfnamefont{E.~J.}\ \bibnamefont{{Wollack}}},\ }%
  \bibfield{journal}{%
  \Doi{10.1088/1475-7516/2011/07/025}{\bibinfo {journal} {J. Cosmol. Astropart.
  Phys.}}\ }%
  \textbf{\bibinfo {volume} {7}},\ \bibinfo {eid} {025} (\bibinfo {year}
  {2011}),\ \Eprint{http://arxiv.org/abs/1105.2044}{arXiv:1105.2044}%
  \bibAnnoteFile{NoStop}{PIXIE}%
\end{thebibliography}%
\end{document}